\pdfoutput=1

\documentclass[fleqn,usenatbib]{mnras}

\usepackage[T1]{fontenc}            
\usepackage{ae,aecompl}
\usepackage{graphics,epsf}
\usepackage{amsmath}                
\usepackage{amsfonts}               
\usepackage{amssymb}                
\usepackage{upgreek}
\usepackage{mathtools}              
\usepackage{epsfig}
\usepackage{epstopdf}
\usepackage{multirow}
\usepackage[para,online,flushleft]{threeparttable}
\usepackage{flushend}
\usepackage{caption, subcaption}    
\usepackage{comment}
\usepackage{booktabs}               
\usepackage{lastpage}               
\usepackage{xhfill}                 

\newcommand{\orcid}[1]{\href{https://orcid.org/#1}{\def\svgwidth{10pt} 
\begingroup%
  \makeatletter%
  \providecommand\color[2][]{%
    \errmessage{(Inkscape) Color is used for the text in Inkscape, but the package 'color.sty' is not loaded}%
    \renewcommand\color[2][]{}%
  }%
  \providecommand\transparent[1]{%
    \errmessage{(Inkscape) Transparency is used (non-zero) for the text in Inkscape, but the package 'transparent.sty' is not loaded}%
    \renewcommand\transparent[1]{}%
  }%
  \providecommand\rotatebox[2]{#2}%
  \newcommand*\fsize{\dimexpr\f@size pt\relax}%
  \newcommand*\lineheight[1]{\fontsize{\fsize}{#1\fsize}\selectfont}%
  \ifx\svgwidth\undefined%
    \setlength{\unitlength}{54bp}%
    \ifx\svgscale\undefined%
      \relax%
    \else%
      \setlength{\unitlength}{\unitlength * \real{\svgscale}}%
    \fi%
  \else%
    \setlength{\unitlength}{\svgwidth}%
  \fi%
  \global\let\svgwidth\undefined%
  \global\let\svgscale\undefined%
  \makeatother%
  \begin{picture}(1,1)%
    \lineheight{1}%
    \setlength\tabcolsep{0pt}%
    \put(0,0){\includegraphics[width=\unitlength,page=1]{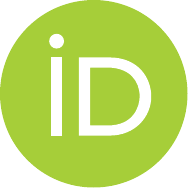}}%
  \end{picture}%
\endgroup%
}}

\DeclareRobustCommand{\ion}[2]{%
\relax\ifmmode
\ifx\testbx\f@series
{\mathbf{#1\,\mathsc{#2}}}\else
{\mathrm{#1\,\mathsc{#2}}}\fi
\else\textup{#1\,{\mdseries\textsc{#2}}}%
\fi}

\usepackage{placeins}

\usepackage{xcolor}
\definecolor{redak}{rgb}{0.9,0.15,0.05}

\title[Optimal CEE $\alpha$ for merging neutron stars]{An optimal envelope ejection efficiency for merging neutron stars}
\author[Tanaka, Gilkis, Izzard \& Tout]{Alexander~M. Tanaka$^{\orcid{0000-0001-5769-5497}}$,$^{1}$\thanks{Present address:
Mathematical Institute, University of Oxford, Woodstock Road, Oxford, OX2 6GG, United Kingdom.}\thanks{E-mail: [firstname].[lastname]@merton.ox.ac.uk (AMT);\newline [firstinitial][lastname]@tauex.tau.ac.il (AG); \newline [firstinitial].[lastname]@surrey.ac.uk (RGI);\newline [initials]@ast.cam.ac.uk (CAT).}
Avishai Gilkis$^{\orcid{0000-0001-8949-5131}}$,$^{2}$
Robert~G. Izzard$^{\orcid{0000-0003-0378-4843}}$ $^{3}$ and Christopher~A. Tout$^{\orcid{0000-0002-1556-9449}}$ $^{1}$
\\
$^{1}$ Institute of Astronomy, University of Cambridge, Madingley Road, Cambridge, CB3 0HA, United Kingdom\\
$^{2}$ The School of Physics and Astronomy, Tel Aviv University, Tel Aviv 6997801, Israel\\
$^{3}$ Astrophysics group, Department of Physics, University of Surrey, Guildford, GU2 7XH, United Kingdom\\
}

\date{Accepted 29 March 2023. Received 29 March 2023; in original form 09 November 2022}

\pubyear{2023}

\volume{522}

\begin{document}
\label{firstpage}
\pagerange{\pageref{firstpage}--\pageref{LastPage}
}

\maketitle

\begin{abstract}
We use the rapid binary stellar evolution code \textsc{binary\_c} to estimate the rate of merging neutron stars with numerous combinations of envelope ejection efficiency and natal kick dispersion. We find a peak in the local rate of merging neutron stars around~$\alpha \approx 0.3$--$0.4$, depending on the metallicity, where $\alpha$ is the efficiency of utilising orbital energy to unbind the envelope. The peak height decreases with increasing electron-capture supernova kick dispersion~$\sigma_\mathrm{ECSN}$. We explain the peak as a competition between the total number of systems that survive the common-envelope phase increasing with~$\alpha$ and their separation, which increases with~$\alpha$ as well. Increasing~$\alpha$ reduces the fraction of systems that merge within a time shorter than the age of the Universe and results in different mass distributions for merging and non-merging double neutron stars. This offers a possible explanation for the discrepancy between the Galactic double neutron star mass distribution and the observed massive merging neutron star event~GW$190425$. Within the $\alpha$--$\sigma_\mathrm{ECSN}$~parameter space that we investigate, the rate of merging neutron stars spans several orders of magnitude up to more than~$1\times 10^{3} \, \mathrm{Gpc}^{-3}\,\mathrm{yr}^{-1}$ and can be higher than the observed upper limit or lower than the observed lower limit inferred thus far from merging neutron stars detected by gravitational waves. Our results stress the importance of common-envelope physics for the quantitative prediction and interpretation of merging binary neutron star events in this new age of gravitational wave astronomy.
\end{abstract}


\section{INTRODUCTION}
\label{sec:intro}

Two merging neutron stars (NSs) were detected for the first time by gravitational-wave emission \citep{GW170817_discovery}, opening a new window into understanding the physics and evolution of compact stars. The simultaneous detection of an electromagnetic counterpart confirmed earlier hypotheses that short-duration gamma-ray bursts \citep{ShortGRBs} and kilonovae \citep{kilonova_term_introduced} result from merging NSs. The synthesis of heavy elements in these merging NSs has also become a topic of great interest \citep{rprocess2017,rprocess2018,Kobayashi2022}.

Besides the implications mentioned above, merging NSs are also related to the evolution of massive stars in binary systems \citep{DeMarcoIzzard2017}. The short orbital period required for two NSs to merge within a time shorter than the age of the Universe suggests that the evolution may have included a common-envelope phase in which the first NS entered the envelope of the evolved companion.

The common-envelope evolution (CEE) phase \citep{CE_paczynski} is not fully understood \citep{CEE2013,CEE2019} and is often characterised by an efficiency parameter~$\alpha$ \citep{CE_webb,CE_livio_1988,CEEalpha2011}. This parameter determines the efficiency of utilising energy released from the decaying orbit inside the common envelope until the removal of the envelope gas itself. In some cases, $\alpha$ can exceed~$1$, representing additional energy sources besides the orbital energy. These include energy release from the recombination of hydrogen \citep*{Han1994} or from accretion on to the dense companion or the core itself \citep{CEEaccretion2017,CEEaccretion2018}. Detailed one-dimensional modelling of a NS entering a CEE phase with a red supergiant companion by~\cite{CEE5} suggests~$\alpha \approx 5$, though this estimate is for envelope removal that continues through post-CEE mass transfer, when the system has become semi-detached once again. However, \cite{alphaScherbakFuller} find~$0.2 \la \alpha \la 0.4$ appropriate for explaining the formation history of white dwarf binaries.

The CEE phase can also affect the formation of radio pulsars: Galactic double NS (DNS) systems, like the Hulse-Taylor binary \citep{HulseTaylor1975}, that are the counterparts of the distant sources of gravitational waves. Even limited mass accretion by a NS during the CEE phase can spin it up \citep{review_1_macleod}, producing a `recycled' pulsar. According to \cite{review_1_chattopadhyay}, NS spin-up and accretion during CEE are important for reproducing the Galactic DNS population.

\begin{figure}
\includegraphics[width=\linewidth]{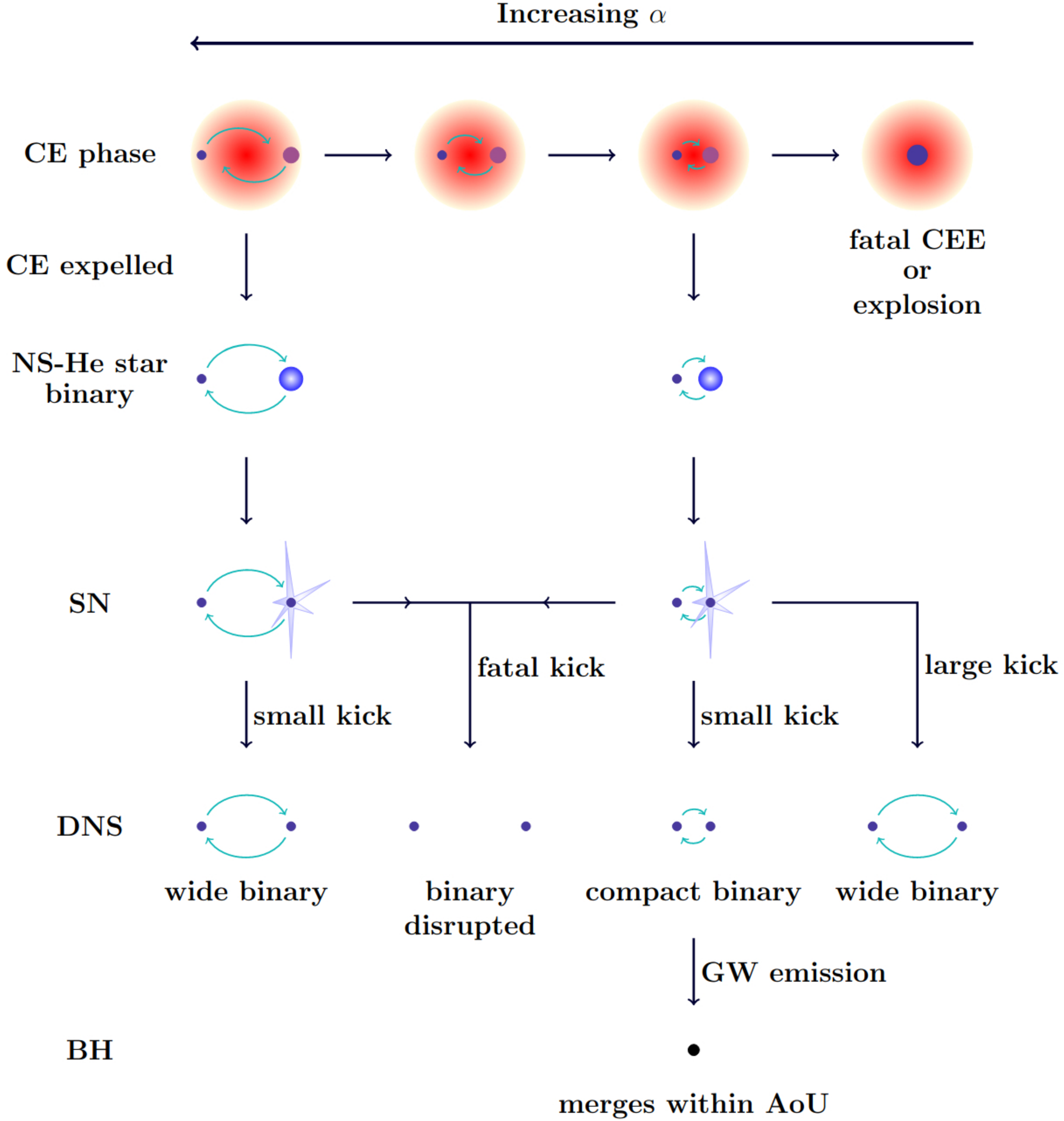}
\caption{The evolution of a double neutron star (DNS) system involving a CEE phase with a NS. The diagram is simplified because the evolution of a DNS may involve multiple CEE phases.}
\label{fig:evol}
\end{figure}

In this paper we address the effect of the uncertainty in the physics of CEE and ejection on the rate of merging of DNSs. The effects we study are depicted schematically in Fig.~\ref{fig:evol} and discussed in detail throughout the paper. In Sect.~\ref{sec:method} we describe the numerical method and model assumptions and in Sect.~\ref{sec:results} we present our key findings. In Sect.~\ref{sec:discussion} we discuss our findings in comparison to previous works and we conclude in Sect.~\ref{sec:conclusions}.

\section{METHOD}
\label{sec:method}

We use version~2.0 of the \textsc{binary\_c} rapid stellar evolution code \citep{binaryc1,binaryc2,binaryc3,binaryc4,IzzardJermyn} to generate synthetic populations of binary systems in isolation. The code is based upon the Binary Star Evolution (\texttt{bse}) code \citep*{binaryc5} with the addition of nucleosynthesis and other changes described in the aforementioned papers. In the following subsections we describe the key details of our modelling.

\subsection{Common-envelope ejection}
\label{subsec:CEE}

We use the common-envelope (CE) $\alpha$-prescription \citep{CE_webb} as described by~\citet{binaryc5}. Following the conventions of~\citet{CE_webb_2008}, we have,
\begin{equation}
    \alpha \bigg(\! \! -\frac{G M_{1\mathrm{c}} M_{2}}{2A_{\mathrm{f}}} + \frac{G M_{1} M_{2}}{2A_{\mathrm{i}}}\bigg) = -\frac{G M_{1} M_{1\mathrm{e}}}{\lambda R_{1\mathrm{,L}}},
\end{equation}
where $M_1$ and~$M_2$ are the binary component masses, $M_{1\mathrm{c}}$ and~$M_{1\mathrm{e}}$ are the core and envelope masses of star~$1$, which is defined to be the star that fills its Roche lobe to initiate the CE,~$A_{\mathrm{i}}$ and~$A_{\mathrm{f}}$ are the semi-major axes of the orbit before and after CEE,~$R_{1 \mathrm{,L} }$ is the Roche-lobe radius of star~$1$ at the start of mass transfer,~$\alpha$ is the dimensionless parameter that encodes the efficiency of converting the binary orbital energy into expelling the CE and~$\lambda$ is the dimensionless parameter scaling the binding energy of different structures of stellar envelopes \citep*{lambda_introduction}. We use the $\lambda$-fit described by~\citet{Claeysetal2014} for the ejection of hydrogen envelopes. Although it was derived at solar metallicity, we use the \citet{Claeysetal2014}~prescription for all our choices of metallicity. \cite{binding_energy_xu_li} and  \cite*{binding_energy_loveridge} derived fitting formulae for additional metallicites. While we do not use these fits, we do test the sensitivity to the envelope binding energy parameterisation by running additional models with $\lambda=0.5$ instead of the \citet{Claeysetal2014} prescription. The $\lambda$-fitting formulae derived by \cite{binding_energy_klencki} are implemented in later \textsc{binary\_c} versions than the one we use here, and we will check how their metallicity dependence affects merging DNSs in a future study.

For CE ejections involving evolved helium stars we resort to simply~$\lambda=0.5$.

\subsection{Maximal NS mass}
\label{subsec:NSmass}

Theoretical calculations of the maximum spherical non-rotating NS mass vary between~$1.3$ and~$2.5\,\mathrm{M}_\odot$ \citep{nsmass_theory} depending on the assumed equation of state for the NS matter. The maximum mass of rigidly or differentially rotating NSs can be significantly higher (\citealt*{review_2.2_baumgarte}; \citealt{nsmass_theory}), and therefore also affected by spin-up during CEE \citep{review_1_chattopadhyay}. However, such hypermassive NSs might not be relevant if the dissipation of differential rotation happens on a short timescale \citep{Duez2006}. Estimates based on observational data from millisecond pulsars and gravitational waves suggest limits on the maximum NS mass between~$2.01$ and~$2.16 \, \mathrm{M}_\odot$ \citep{nsmass1,nsmass2}, though the analysis by \cite{mass_GW_inference_landry} suggests that the population of NSs observed in gravitational waves is consistent with a maximal NS mass of $2.7\,\mathrm{M}_\odot$. We employ a conservative bound of~$2.0 \, \mathrm{M}_\odot$ as the maximum NS mass in our simulations.

\subsection{Natal kicks}
\label{subsec:NSkick}

NSs produced by supernovae (SNe) experience natal kicks because of the asymmetry of the explosion \citep{LyneLorimer1994}. There is no consensus about the fundamental cause of these kicks but it has been suggested that they result from spherical accretion-shock instabilities that occur in the post-bounce phase of core-collapse supernovae \citep{accretion_instability_Blondin_2006} or convective instabilities of the neutrino heated layer in the core \citep{kick_neutrino_scheck}.

In our simulations, all the natal kick speeds are sampled from a singly-peaked Maxwellian probability density function~$\varphi_v$,
\begin{gather}
\varphi_v \, \mathrm{d}v = \sqrt{\frac{2}{\pi}} \frac{v^2}{\sigma^3} e^{-v^2 / 2\sigma^2} \mathrm{d}v \: , \label{max_kick}
\end{gather}
where $v$ is the natal kick speed and~$\sigma$ is the natal kick speed dispersion that depends upon the type of supernova. The direction of the natal kick is randomly chosen over a sphere. \cite{hobbs} found in a study of~$233$ pulsar proper motions that the distribution of pulsar natal speeds obeys a singly-peaked Maxwellian distribution with~$\sigma = 265 \, \mathrm{km}\, \mathrm{s}^{-1}$.

Alternatively, it has been inferred that NS natal kick speeds are more accurately described by a bimodal distribution than a singly peaked Maxwellian \citep*{bimodal_kick_Fryer_1998, bimodal_kick} because of the differing nature of the various SN origins of the NSs. Electron-capture supernovae \citep*[ECSNe;][]{ecsne_first} might have slower natal kicks \citep{ecsne_low_kick_2} because they explode too quickly for convective instabilities to form. ECSNe are also more likely to occur in close binaries \citep{ecsne_low_kick, Poelarendsetal2017}.

We use a natal-kick speed dispersion of~$\sigma=265\,\mathrm{km}\,\mathrm{s}^{-1}$ for Type Ib/c and Type II supernovae. We vary the natal-kick speed of ECSNe such that~$\sigma_{\mathrm{ECSN}} \in$ \{$0$, $7$, $15$, $26$, $265\} \, \mathrm{km}\, \mathrm{s}^{-1}$, similarly to \citet{giacobbo_ecsne_dns}. The differing natal-kick speeds of different SN types in our simulations effectively create an overall bimodal natal kick distribution in NSs.

\subsection{Time to merge by gravitational wave emission}
\label{subsec:GWdelay}

The delay time~$t$ of a DNS is the time difference between the zero-age main sequence (ZAMS) phase of the binary and merging. DNSs merge owing to the loss of angular momentum from gravitational wave radiation. The time to merge for two point masses moving in an elliptical orbit can be approximated by using an expansion of the field equations of general relativity \citep{GW_rad}. We implement a computationally efficient numerical fit (Appendix~\ref{sec:appendixa}) to this approximation to calculate the time~$T$ it takes the DNS to merge,
\begin{multline} 
    T(a_0,e_0) =  \frac{5}{256} \frac{c^5}{G^3} \frac{a_{0}^{4}}{m_1 m_2 (m_1 + m_2)} \\
    \times \Bigg[\frac{768}{425} + \displaystyle\sum\limits_{i=1}^{10}\beta_{i}\big(1 - e_{0}^2 \big) ^ {i/2} \Bigg] \big(1 - e_{0}^2 \big) ^{7/2} \: , \label{eqn:numerical_fit_method}
\end{multline}
where $a_0$ is the semi-major axis of the binary orbit when the DNS forms,~$e_0$ is the eccentricity of the orbit,~$m_{1}$ and~$m_{2}$ are the masses of the binary components and~$\beta_i$ are given in Table~\ref{tab:constants} in Appendix~\ref{sec:appendixa}. We do not account for the effect of the first-order post-Newtonian perturbation on the gravitational wave induced decay time-scale as outlined by \cite{Zwicketal2020}.
\begin{figure*}
  \centering
    \begin{subfigure}{0.31\linewidth}
    \centering
    \includegraphics[width=\linewidth]{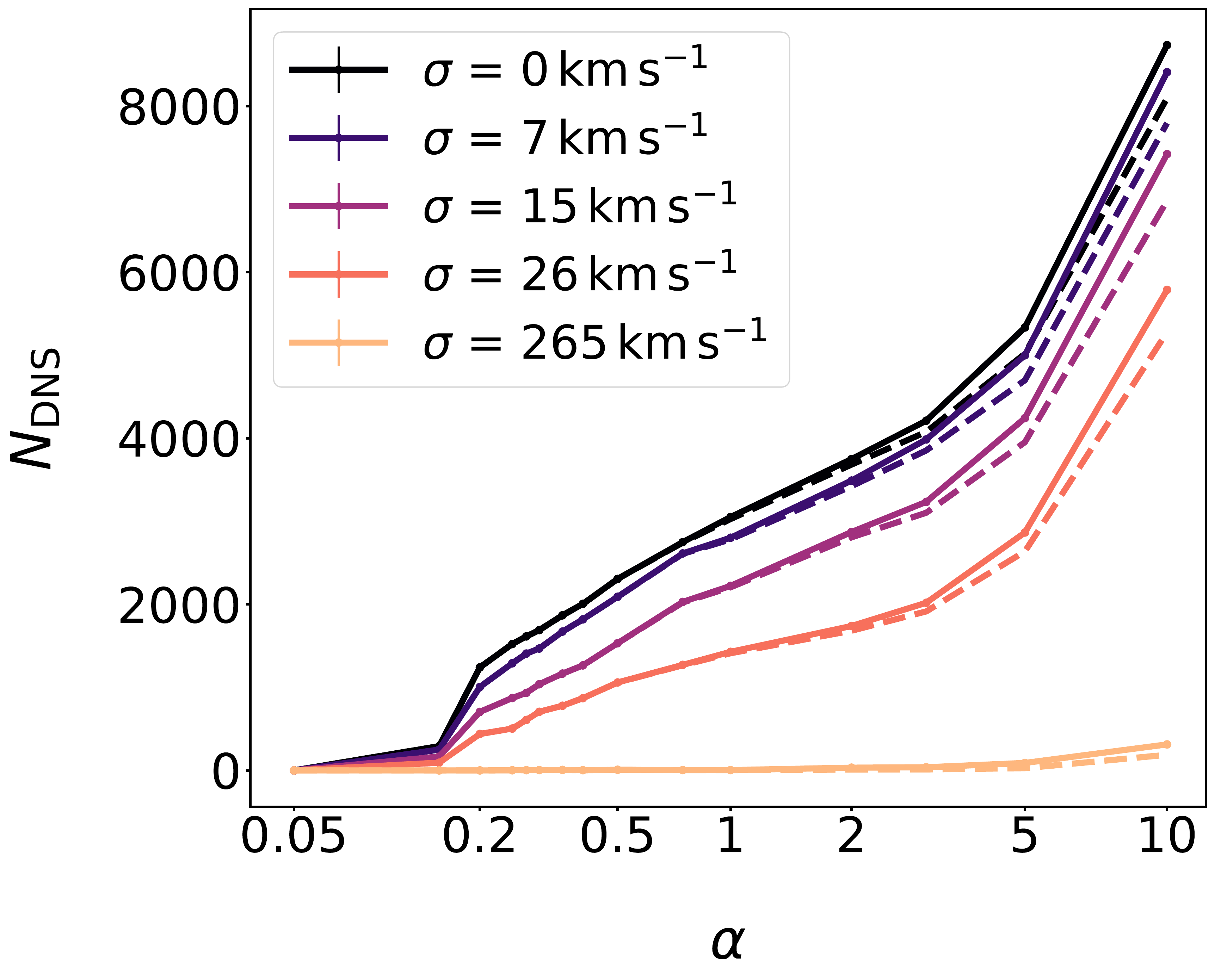}
    \caption{$Z = 0.02$.}\label{fig:all_m02}
  \end{subfigure}
    \hspace{2ex}
  \begin{subfigure}{0.31\linewidth}
    \centering
    \includegraphics[width=\linewidth]{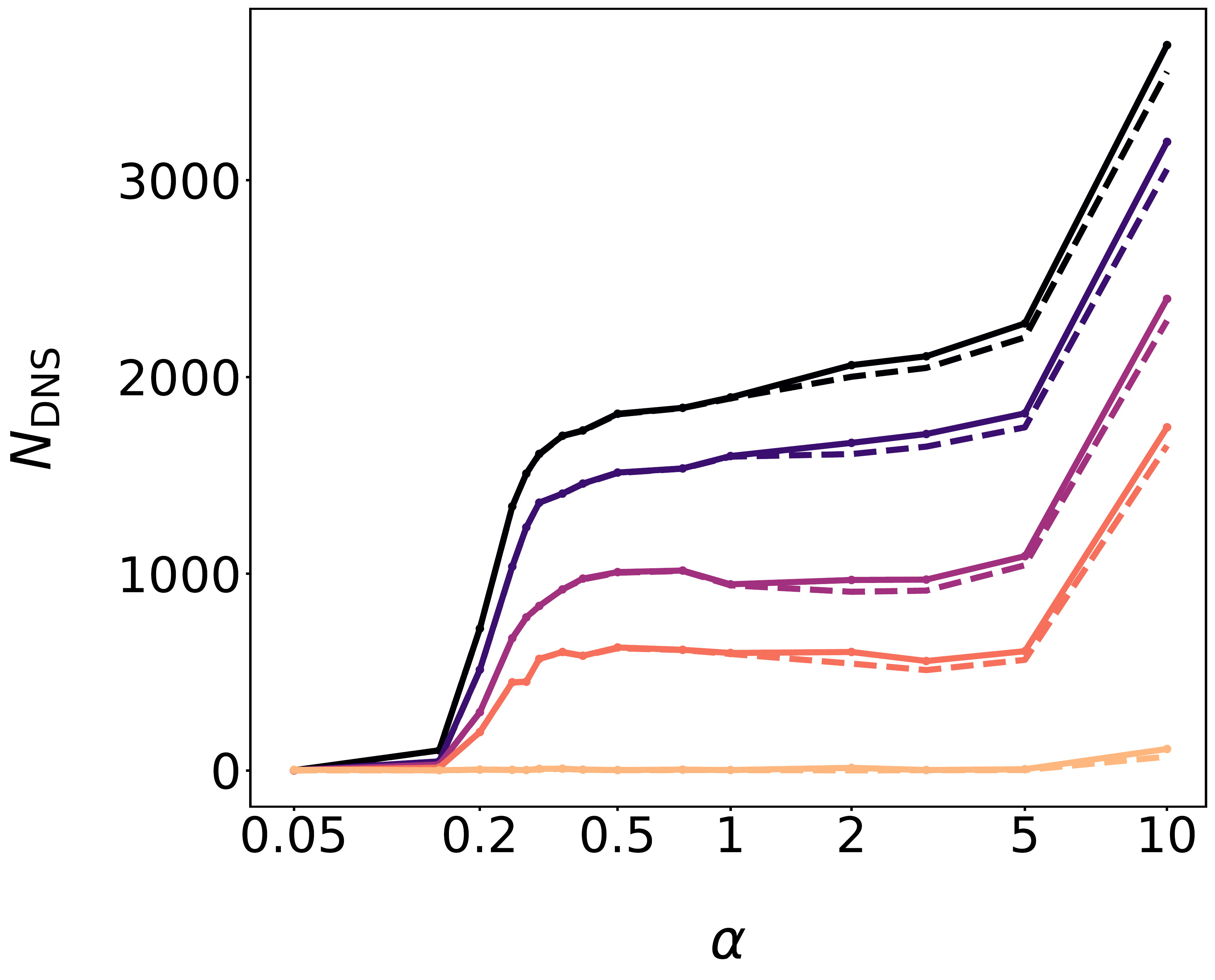}
    \caption{$Z = 0.002$.}\label{fig:all_m002}
  \end{subfigure}
      \hspace{2ex}
    \begin{subfigure}{0.31\linewidth}
    \centering
    \includegraphics[width=\linewidth]{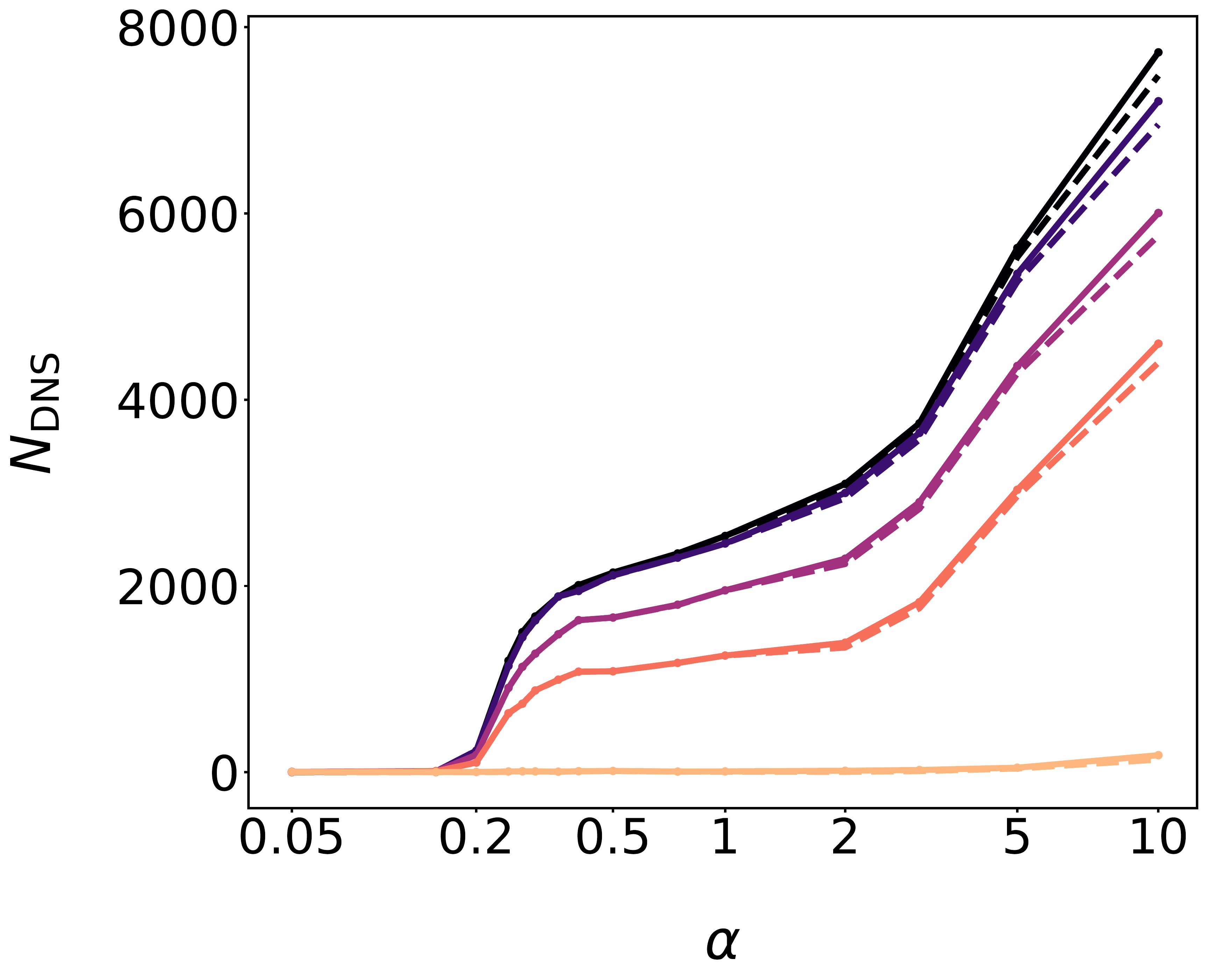}
    \caption{$Z = 0.0002$.}\label{fig:all_m0002}
  \end{subfigure}\\
  \vspace{1ex}
  \centering
    \begin{subfigure}{0.31\linewidth}
    \centering
    \includegraphics[width=\linewidth]{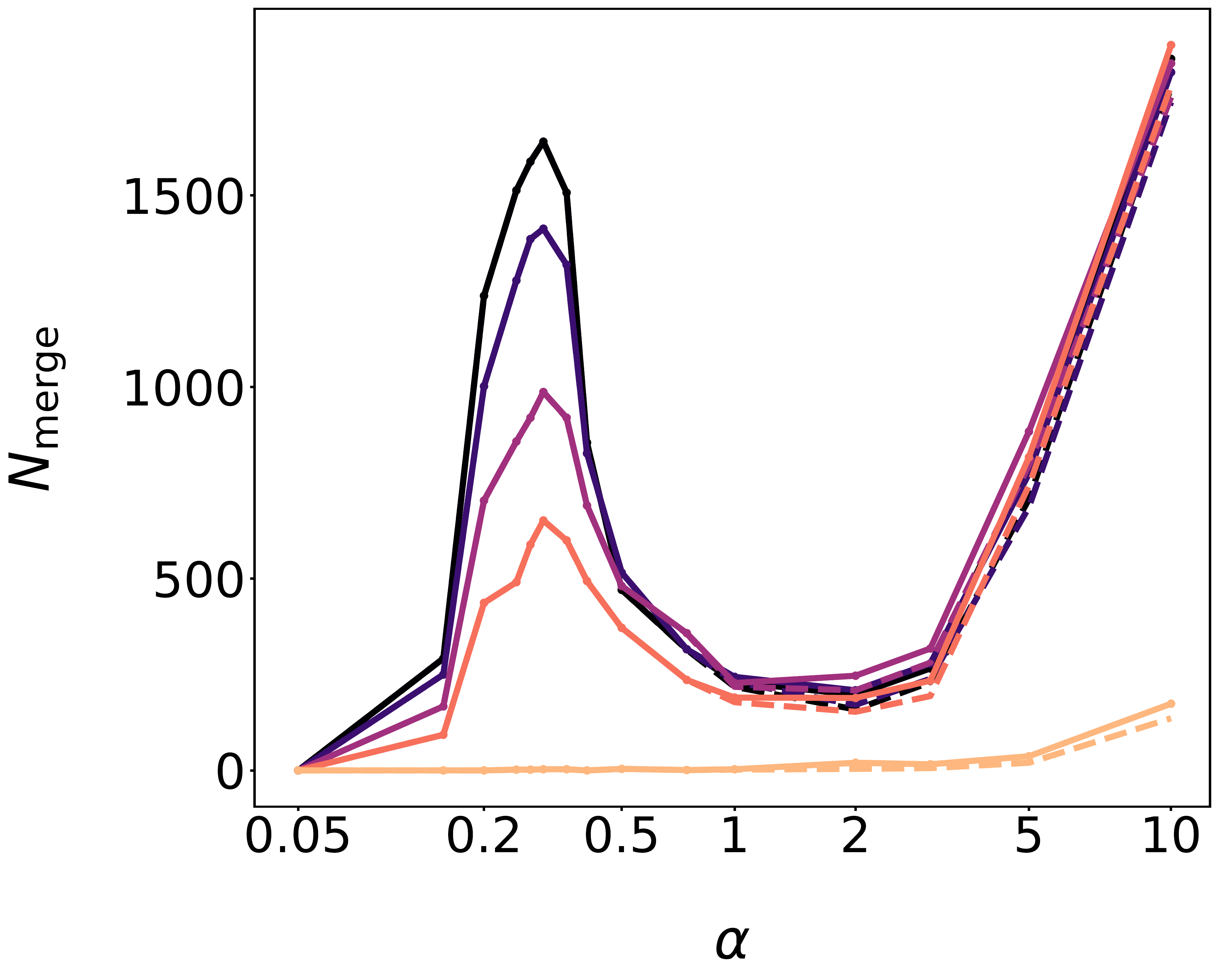}
    \caption{$Z = 0.02$.}\label{fig:merge_m02}
  \end{subfigure}
    \hspace{2ex}
  \begin{subfigure}{0.31\linewidth}
    \centering
    \includegraphics[width=\linewidth]{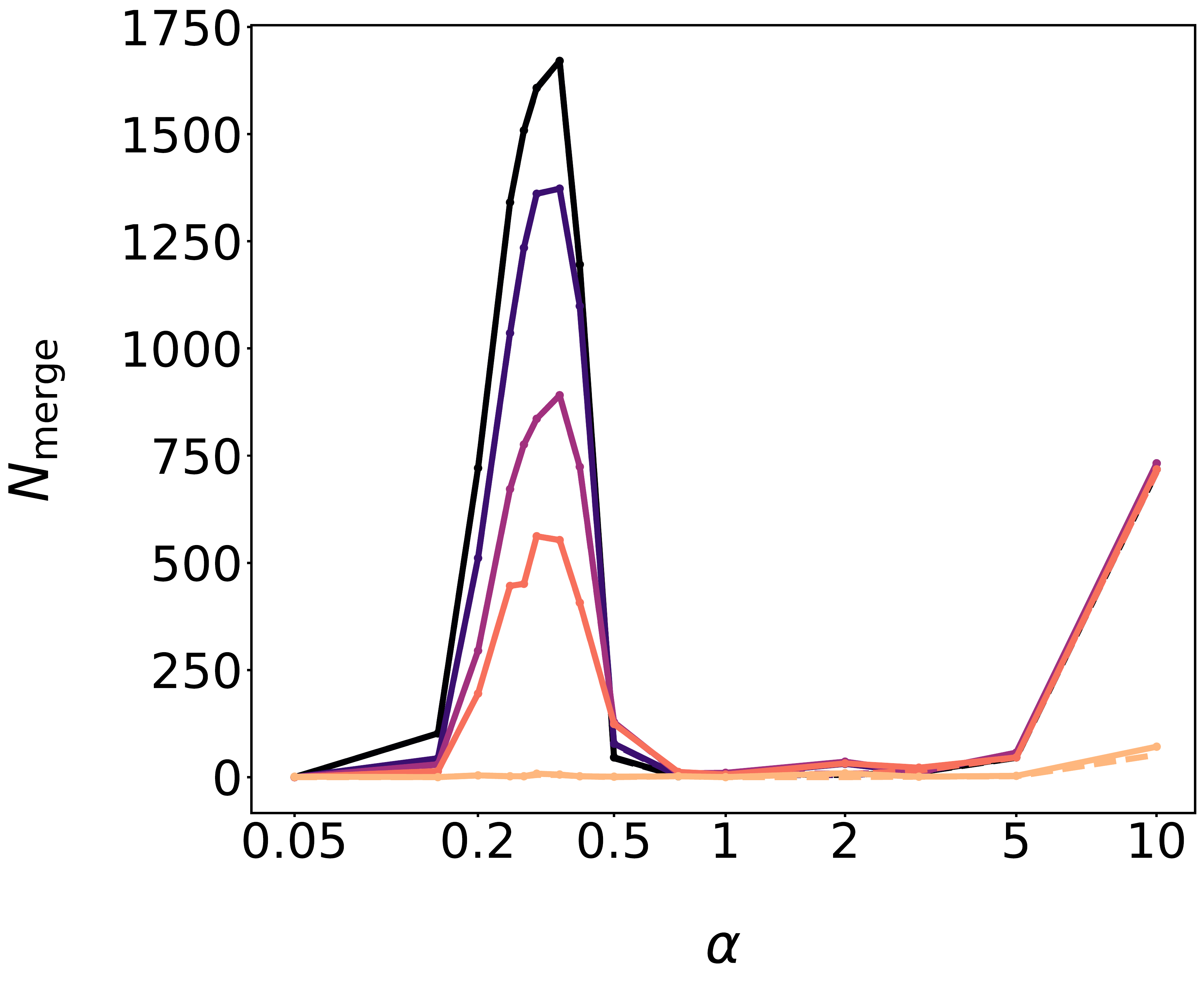}
    \caption{$Z = 0.002$.}\label{fig:merge_m002}
  \end{subfigure}
      \hspace{2ex}
    \begin{subfigure}{0.31\linewidth}
    \centering
    \includegraphics[width=\linewidth]{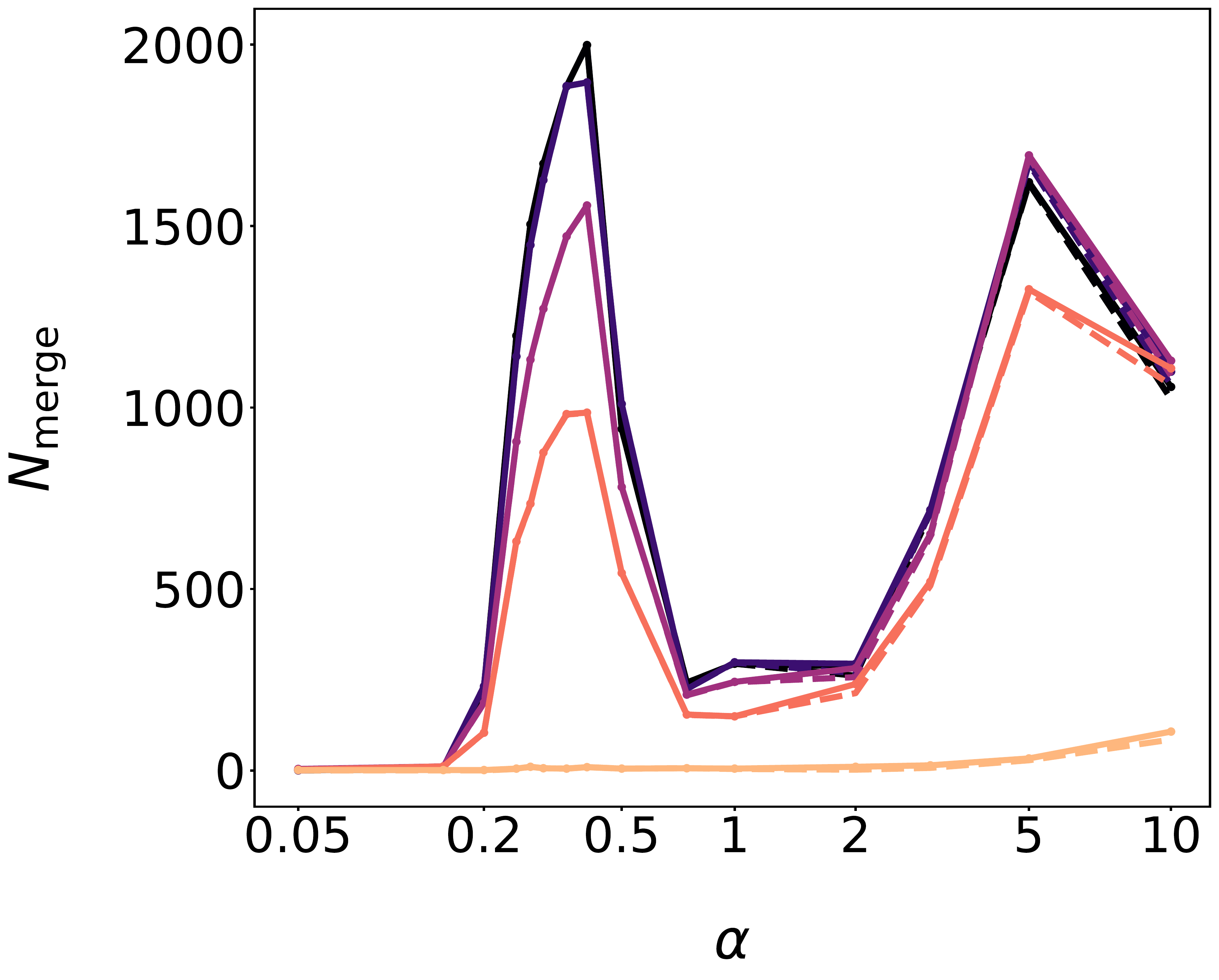}
    \caption{$Z = 0.0002$.}\label{fig:merge_m0002}
  \end{subfigure}\\  
  \caption{A comparison of different models with ECSN kick speed dispersions of~$\sigma_\mathrm{ECSN}\in \{0$, $7$, $15$, $26$, $265\} \, \textnormal{km}\,{\textnormal{s}}^{-1}$ at metallicities of~$Z \in \{0.02$, $0.002$, $0.0002$\}. Panels~\subref{fig:all_m02}--\subref{fig:all_m0002} display the total number of DNS systems formed, $N_\mathrm{DNS}$, as function of~$\alpha$. Panels~\subref{fig:merge_m02}--\subref{fig:merge_m0002} display the number~$N_\mathrm{merge}$ of DNS systems that merge within the age of the Universe ($13.8 \, \mathrm{Gyr}$) as function of~$\alpha$. The dashed lines show the binaries that went through at least one CEE phase involving a NS during their evolution. These are included in the total number presented by the solid lines.}
  \label{fig:all_m}
\end{figure*}

\subsection{Grid parameters}
\label{subsec:grid}

The limits of the simulation grid parameter space~$\widetilde{\Omega}$ were chosen by initially running a sparser grid to ascertain which area of the parameter space leads to DNS formation. This allowed us to increase the effective resolution of our denser grids without having to increase its size. The masses of the primary star~$M_1$, defined to be the initially more massive of the two stars, are distributed uniformly in log-space between~$6$ and~$90\, \mathrm{M}_\odot$ at $72$~intervals. The mass ratios~$Q\equiv~M_{2}/M_{1}$, from which the masses of the secondary stars are calculated, are distributed uniformly between~$0.3$ and~$1.0$ at $71$~intervals. The initial periods are distributed uniformly in log-space between~$1$ and~$10\,000 \, \mathrm{d}$ at $72$~intervals.

\subsection{Local DNS merging rate}
\label{subsec:merger_rate}

When computing quantities based on the simulated population, such as the average NS mass, we weight each sampled binary according to its initial parameter probability distributions. We use the probability density functions~$\varphi_X$ and sample spaces~$\Omega_X$ for the continuous random variables~$X$ given by the ZAMS primary masses~$M$, mass ratios~$Q$ and periods~$P$\,:
\begin{gather}
    \varphi_M \, \mathrm{d} M \propto M^{-2.3}\,\mathrm{d} M, \: \Omega_M = [0.08, 150] \, \mathrm{M}_\odot \, , \nonumber \\
    \varphi_Q \, \mathrm{d} Q \propto \mathrm{d} Q \, , \: \Omega_Q  = [0.1, 1.0] \ \text{and} \nonumber \\
    \varphi_{P} \, \mathrm{d} P \propto P^{-1} \mathrm{d} P \, , \: \Omega_{P} = [1, 10\,000] \, \mathrm{d} \, , \label{eqn:pdf_MqP}
\end{gather}
with suitable normalisations to ensure that $\int_{X(\Omega_X)} \varphi_X(x) \, \mathrm{d} x = 1$. Our metallicity probability distribution is a log-normal distribution \citep{metallicity_lognormal_Penprase_2010, metallicity_0.5_Rafelski}, 
\begin{multline}
    \varphi_{\log_{10} Z} \, \mathrm{d} \log_{10} \left(Z / \mathrm{Z}_\odot\right) \\ \propto \exp \Bigg(\! \! - \frac{\big( \log_{10} \left(Z / \mathrm{Z}_\odot\right) - \mu \big) ^2 }{2 \sigma_{\log_{10} Z}^2 } \Bigg) \mathrm{d} \log_{10} \left(Z / \mathrm{Z}_\odot\right) \, ,
\end{multline}
\begin{gather}
    \Omega_{\log Z} = [-2.5, 0.5] \, , \label{eqn:pdf_Z}
\end{gather}
where $Z$ is the metallicity, $\mathrm{Z}_\odot \equiv 0.02$ and the logarithmic metallicity dispersion~$\sigma_{\log_{10} Z} = 0.2$ \citep{Santoliquido_2020} or~$\sigma_{\log_{10} Z} = 0.5$, which are not dependent on redshift \citep{metallicity_0.5_Rafelski, metallicity_dispersion_0.55_Neeleman_2013}. We implement metallicity evolution \citep{metallicity_Savaglio,metallicity_Prochaska_2003} by taking the mean logarithmic metallicity~$\mu$ to be a linear function of the redshift~$z$ \citep{Santoliquido_2020},
\begin{align}
    \mu (z) = \log_{10} \left(a / \mathrm{Z}_\odot\right) + b z \, , \label{eqn:mean_Z}
\end{align}
where $a = 1.04\, \mathrm{Z}_\odot$ and~$b = -0.24$. Similarly to \cite{Santoliquido_2020}, we have chosen~$a$ to be consistent with the averaged stellar metallicity found by \cite{metallicity_average_Gallazi}, based upon the Sloan Digital Sky Survey Data Release~$2$ and~$b$ to be consistent with the slope fitted by \cite{metalllicity_De_Cia}, based upon the mean metallicity (weighted for the \ion{H}{i}  content) from a variety of damped Lyman-$\alpha$ absorber samples up to redshift~$z=5$. 

\begin{figure*}
  \centering
    \begin{subfigure}{0.31\linewidth}
    \centering
    \includegraphics[width=\linewidth]{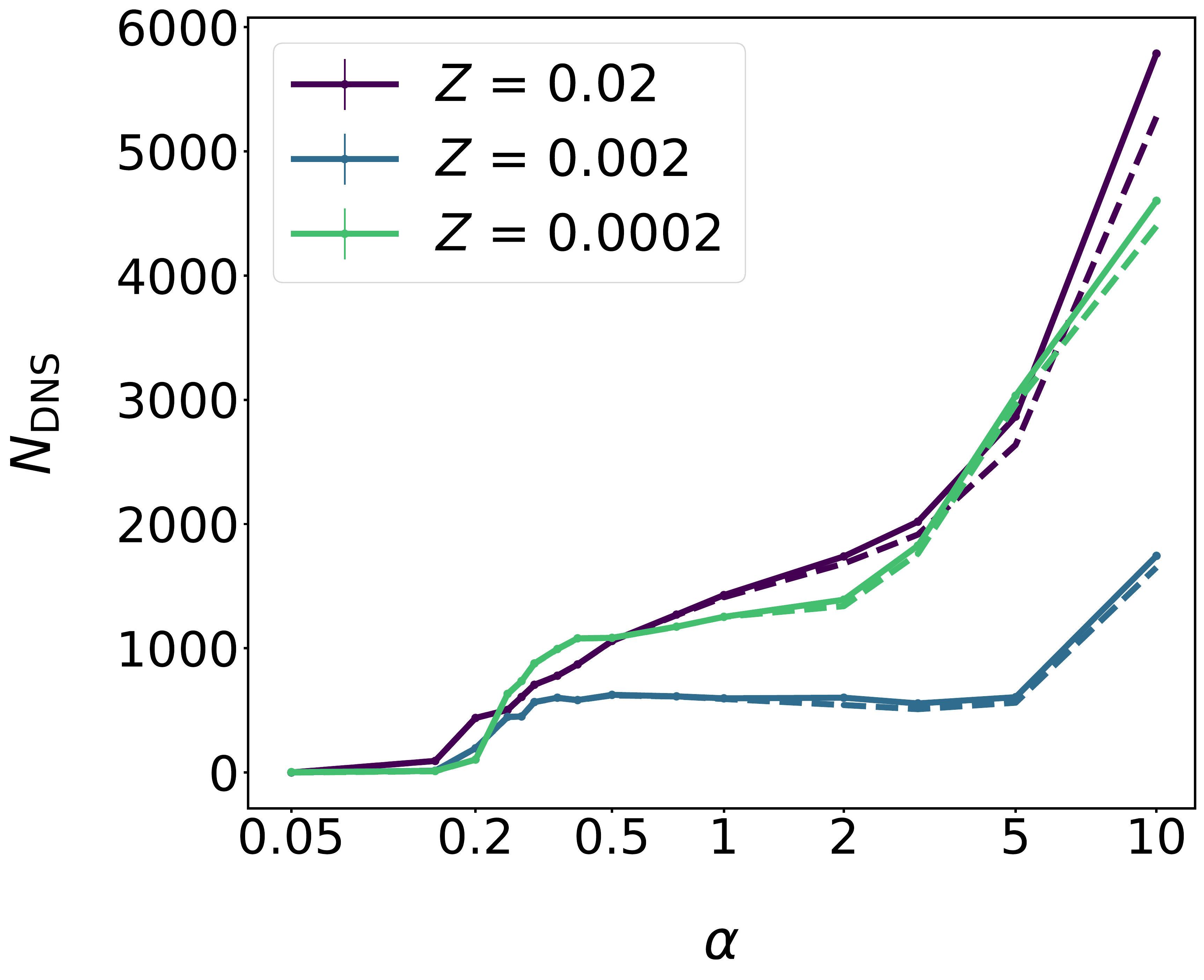}
    \caption{}\label{fig:all_ecap26}
  \end{subfigure}
    \hspace{2ex}
  \begin{subfigure}{0.31\linewidth}
    \centering
    \includegraphics[width=\linewidth]{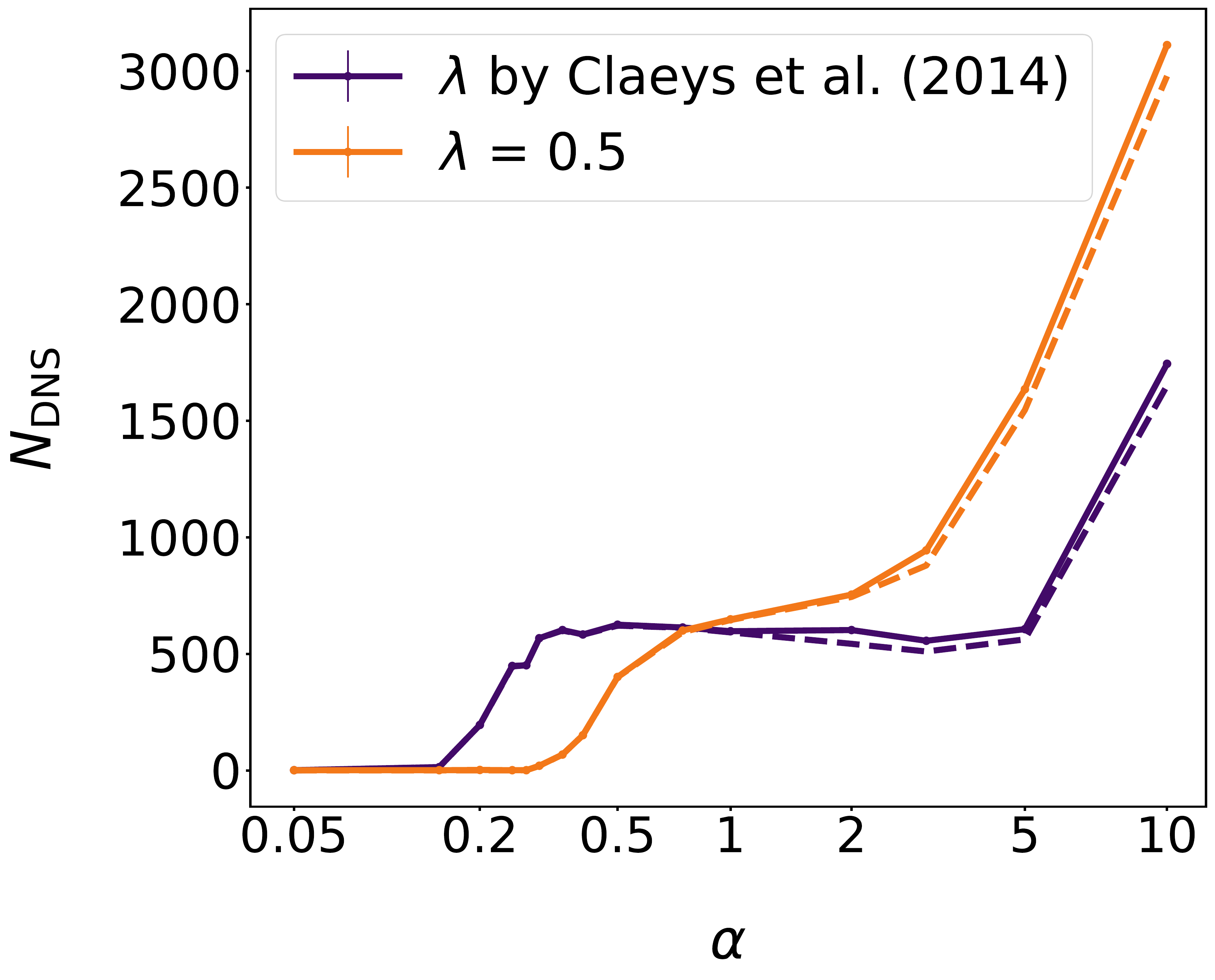}
    \caption{$Z = 0.002$.}\label{fig:form_lambda_002}
  \end{subfigure}
      \hspace{2ex}
    \begin{subfigure}{0.31\linewidth}
    \centering
    \includegraphics[width=\linewidth]{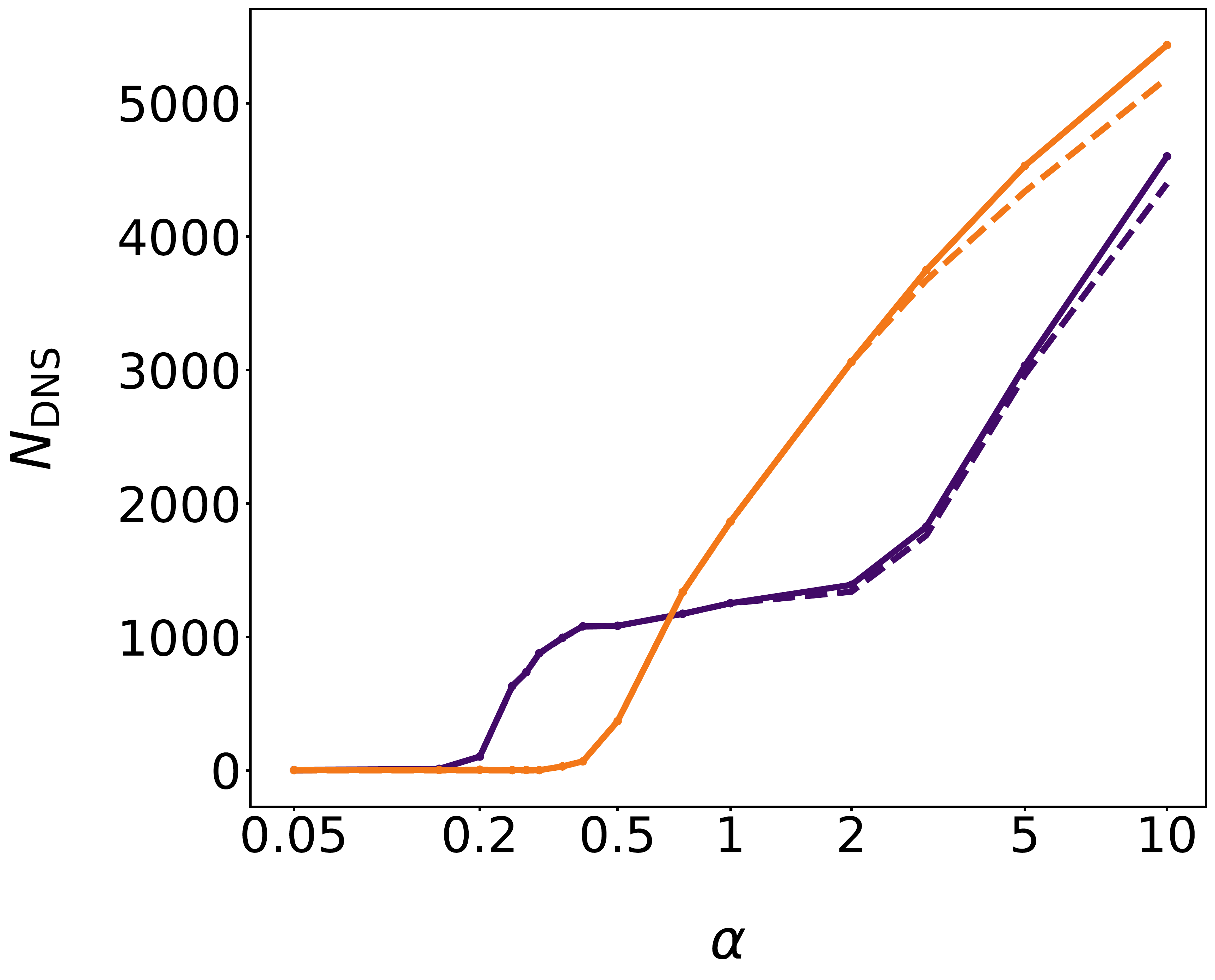}
    \caption{$Z = 0.0002$.}\label{fig:form_lambda_0002}
  \end{subfigure}\\
  \vspace{1ex}
  \centering
    \begin{subfigure}{0.31\linewidth}
    \centering
    \includegraphics[width=\linewidth]{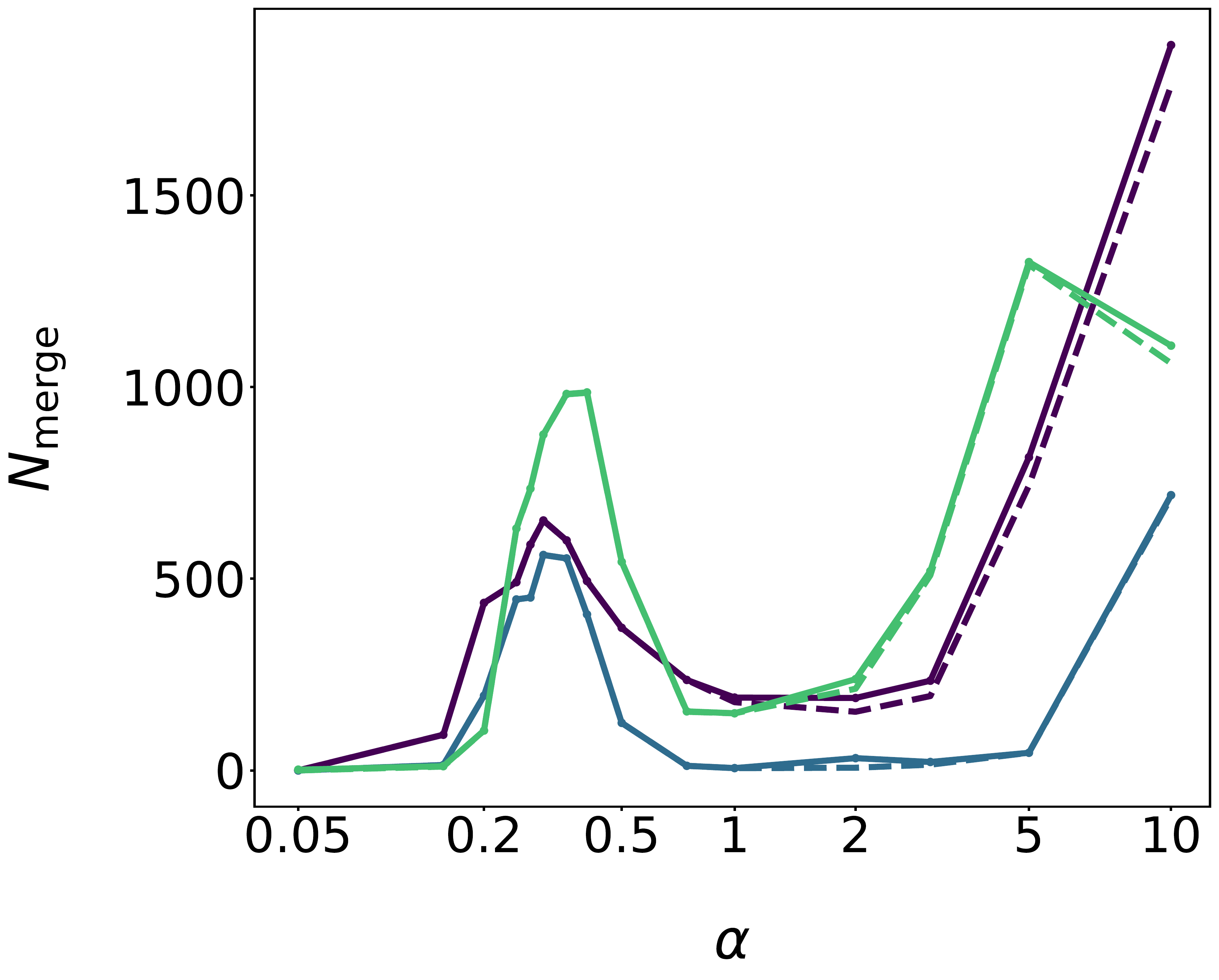}
    \caption{}\label{fig:merge_ecap26}
  \end{subfigure}
    \hspace{2ex}
  \begin{subfigure}{0.31\linewidth}
    \centering
    \includegraphics[width=\linewidth]{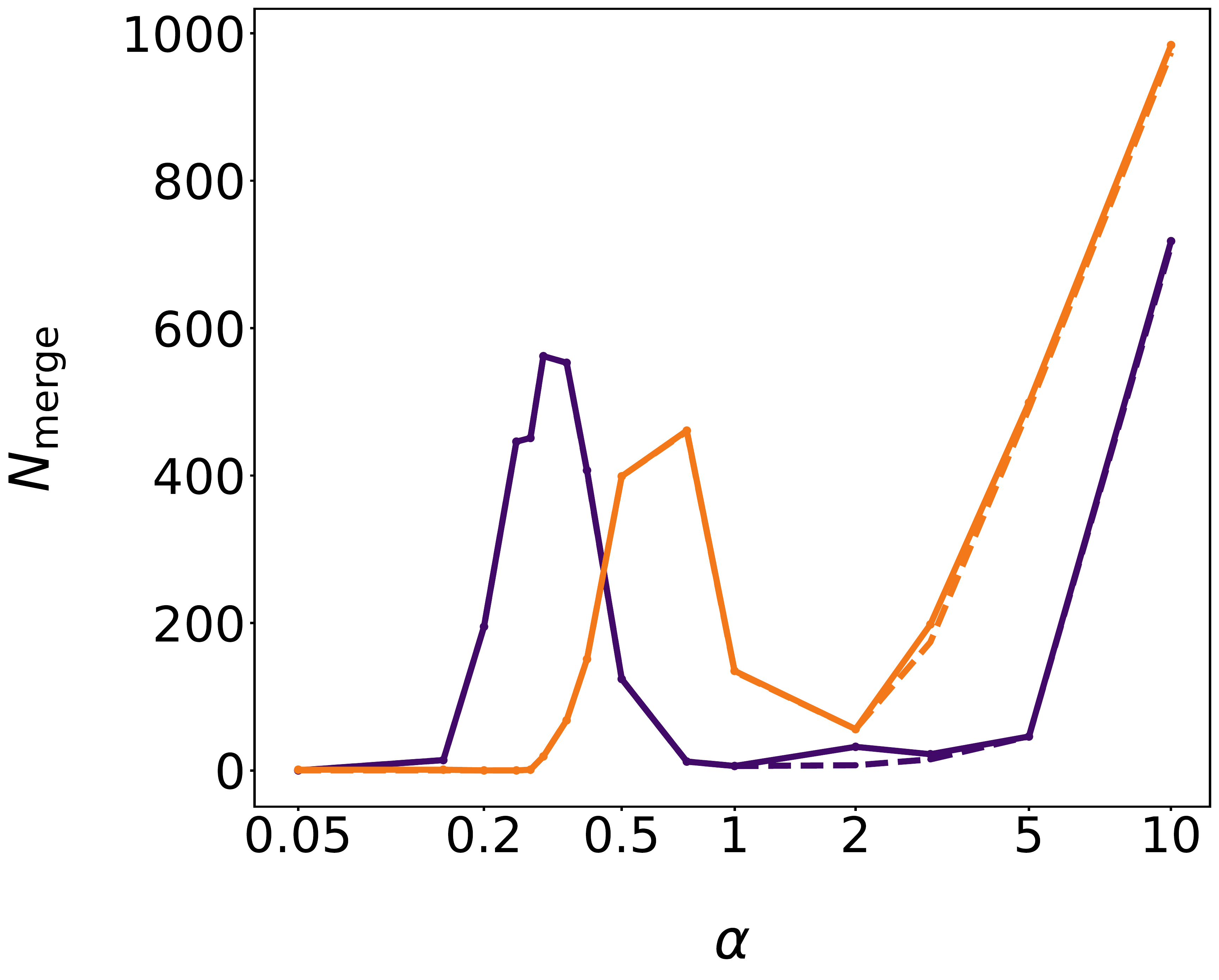}
    \caption{$Z = 0.002$.}\label{fig:merge_lambda_002}
  \end{subfigure}
      \hspace{2ex}
    \begin{subfigure}{0.31\linewidth}
    \centering
    \includegraphics[width=\linewidth]{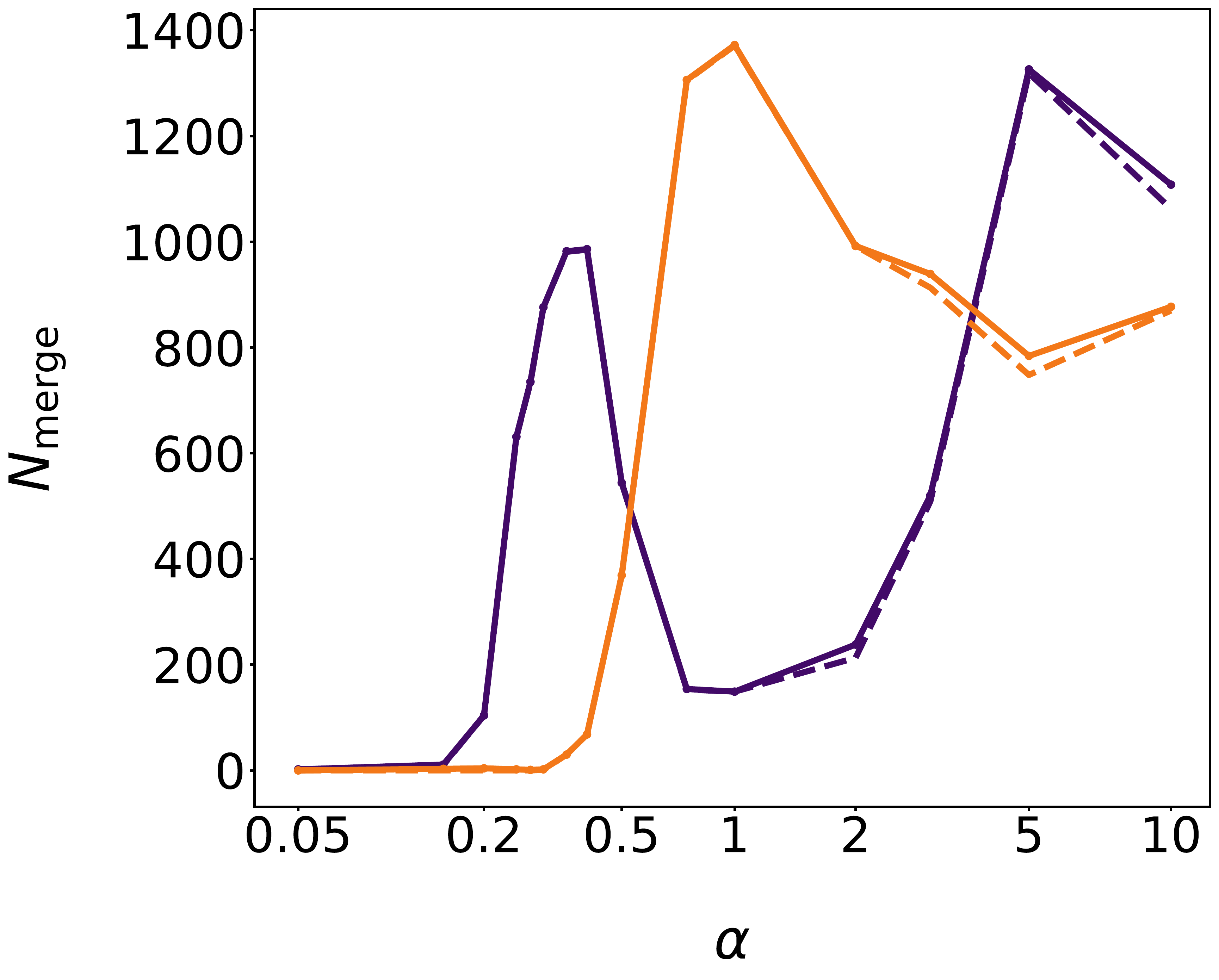}
    \caption{$Z = 0.0002$.}\label{fig:merge_lambda_0002}
  \end{subfigure}\\  
  \caption{The total number of DNS systems (top panels) and the number of merging DNSs (bottom panels) for varying metallicity and $\lambda$-prescriptions. All simulations have the same ECSN kick dispersion of~$\sigma_\mathrm{ECSN}= 26 \, \textnormal{km}\, {\textnormal{s}}^{-1}$. Panels~\subref{fig:all_ecap26} and~\subref{fig:merge_ecap26} are a comparison for the same model assumptions with different metallicities. Comparisons between the~\citet{Claeysetal2014} $\lambda$-prescription (fit to~\citealt{lambda_dewi_2000}) and taking~$\lambda = 0.5$ are shown in Panels~\subref{fig:form_lambda_002} and~\subref{fig:merge_lambda_002} for $Z = 0.002$ and in Panels~\subref{fig:form_lambda_0002} and~\subref{fig:merge_lambda_0002} for~$Z = 0.0002$. The dashed lines show the binaries that went through at least one CEE phase involving a NS during their evolution.}
  \label{fig:m_and_lambda}
\end{figure*}

We adopt the cosmic stellar-formation rate density (SFRD) of~\cite{SFRD_madau}. This is valid up to~$z \approx 8$ and has the form,
\begin{align}
    \mathrm{SFRD}(z) = 0.015 \frac{(1 + z)^{2.7}}{1 + \big[ (1 + z)/2.9 \big]^{5.6}} \, \mathrm{M}_\odot \,  \mathrm{Mpc}^{-3} \, \mathrm{yr}^{-1} \, . \label{eqn:sfrd}
\end{align}
We approximate the local DNS merger rate~$\mathcal{R}$ at the present cosmic time~$t_0$ by averaging the merging rate over some local time interval~$[t_0 -\Delta t, t_0]$ and only considering contributions from DNSs that formed later than a minimum cosmic time~$t_\mathrm{min}$. This gives us,
\begin{multline}
    \mathcal{R}(t_0) \approx \frac{1}{\Delta t}\int_{t_0 - \Delta t}^{t_0} \!\!\!\mathrm{d} t\int_{t_\mathrm{min}}^{t} \!\!\!\mathrm{d} t' \frac{\mathrm{SFRD} \big( z(t') \big) }{\mathbb{E} \big[ M ( 1 + Q ) \big]  } f_\mathrm{bin} \varphi_t( t - t') \, , \label{eqn:conv_av} 
\end{multline}
where $\mathbb{E} \big[M(1 + Q) \big] \approx 0.48 \,\mathrm{M}_\odot$ is the expected total ZAMS mass of a binary, $f_\mathrm{bin}$ is the fraction of ZAMS stars in binaries, $\varphi_t$ is the probability density of DNS delay times~$t$ with respect to the reference measure~$\mathrm{d} t$, $z$ is the redshift, $t_\mathrm{min} = t_0 - t_L(z_\mathrm{max})$ and~$t_L$ is the lookback time as a function of redshift. The size of the local time interval~$\Delta t$ is chosen \textit{a posteriori} to be the minimum uniform bin width of the simulated delay time histograms that preserves the signal-to-noise ratio of the histograms. After integrating to calculate the average, Eq.~\eqref{eqn:conv_av} can be then approximated using a middle Riemann sum to
\begin{multline}
    \mathcal{R}(t_0) \approx f_\mathrm{bin} \frac{\Delta t_L }{\Delta t } \sum_{i = 0}^{N - 1} \frac{ \mathrm{SFRD} \big( z ( t_{i+0.5} ) \big) }{ \mathbb{E} \big[ M ( 1 + Q ) \big] } \\
    \times \mathbb{P} \Big( t \in \big( R ( t_{i+0.5} - \Delta t ), \,t_{i+0.5} \big] \Big) \, ,
\end{multline}
where $\Delta t_L \equiv t_L(z_\mathrm{max}) / N$ is the width of the Riemann sum partitions, $N$ is the number of partitions, $t_{j} \equiv j \Delta t_L$, $\mathbb{P} \big( t \in (t_1, t_2] \big)$ is the probability\footnote{The notion of the probability of an event and the expectation of a function over the grid is formalised in Appendix~\ref{sec:appendixb}.} of a DNS with a delay time~$t$ within the half-open time interval~$(t_1, t_2]$ and $R$ is the ramp function. We use the \texttt{z\_at\_value} and \texttt{Planck15.lookback\_time} functions from the \texttt{cosmology} package of \textsc{astropy} \citep{astropy1, astropy2} to calculate the redshift as a function of the lookback time~$z(t_L)$. We assume a flat $\Lambda$CDM~model and the parameters from the~\cite{planck15}. We use~$z_\mathrm{max} = 8$, $N = 100$, $\Delta t = 1.5 \, \mathrm{Gyr}$ and $f_\mathrm{bin} = 0.7$ (unless explicitly varied; \citealt{binary_fraction}) for our calculations. We do not take into account selection effects, noise, or any artefacts that might result from imperfect measurement because we approximate the astrophysical rate of merging DNSs rather than the realisation of a specific survey.

\section{RESULTS}
\label{sec:results}

We ran~$15$ sets of simulations with varying~$\sigma_{\mathrm{ECSN}} \in \{0$, $7$, $15$, $26$, $265\} \, \mathrm{km}\, \mathrm{s}^{-1}$ and metallicity~$Z \in \{0.02$, $0.002$, $0.0002\}$. Each set of simulations had $\alpha \in \{0.05$, $0.15$, $0.2$, $0.25$, $0.275$, $0.3$, $0.35$, $0.4$, $0.5$, $0.75$, $1$, $2$, $3$, $5$, $10$\} and~$\lambda$ computed by the \cite{Claeysetal2014}~prescription. Two further sets of simulations were run with varying metallicity~$Z \in$ \{$0.002$, $0.0002\}$ with a constant binding-energy parameter~$\lambda = 0.5$ and constant~$\sigma_{\mathrm{ECSN}} = 26 \, \mathrm{km}\, \mathrm{s}^{-1}$. In total~$255$ grids were computed (each consisting of~$368\,064$ binary evolution simulations).

\subsection{Number of merging DNSs}
\label{subsec:number_of_merging_DNSs}

In Fig.~\ref{fig:all_m02}--\subref{fig:all_m0002} we show the total number of NS binaries that form and in Fig.~\ref{fig:merge_m02}--\subref{fig:merge_m0002} the number of binaries that merge over a range of metallicities. If the delay time of a DNS is shorter than the age of the Universe,~$13.8\, \mathrm{Gyr}$ \citep{aou_planck} in our simulations, then the DNS is considered to have merged. Increasing~$\alpha$ eases envelope ejection and so more binaries survive the CEE phase leading to monotonic increase in the number of DNSs with~$\alpha$ (Fig.~\ref{fig:all_m02}--\subref{fig:all_m0002}). Decreasing~$\alpha$ requires the investment of more orbital energy and so the post-CEE separations and the time to coalescence are shorter. The combination of both effects produces the local minima and maxima in the number of merging DNSs (Fig.~\ref{fig:merge_m02}--\subref{fig:merge_m0002}). The local maximum in the number of merging NSs occurs at envelope ejection efficiencies between~$\alpha = 0.30$ and~$\alpha = 0.40$, increasing as the metallicity decreases.

Excluding grids with low numbers of DNSs (such as those with~$\sigma_{\mathrm{ECSN}} = 265 \, \textnormal{km}\, {\textnormal{s}}^{-1}$ and~$\alpha = 0.05$), $81$ to~$100$ per cent of DNSs passed through at least one CE with a NS during their evolution. Up to~$0.7$ per cent of all DNSs and up to~$1.7$ per cent of merging DNSs pass through a white dwarf accretion induced collapse to form one of the NSs.

\begin{figure*}
  \centering
    \begin{subfigure}{0.31\linewidth}
    \centering
    \includegraphics[width=\linewidth]{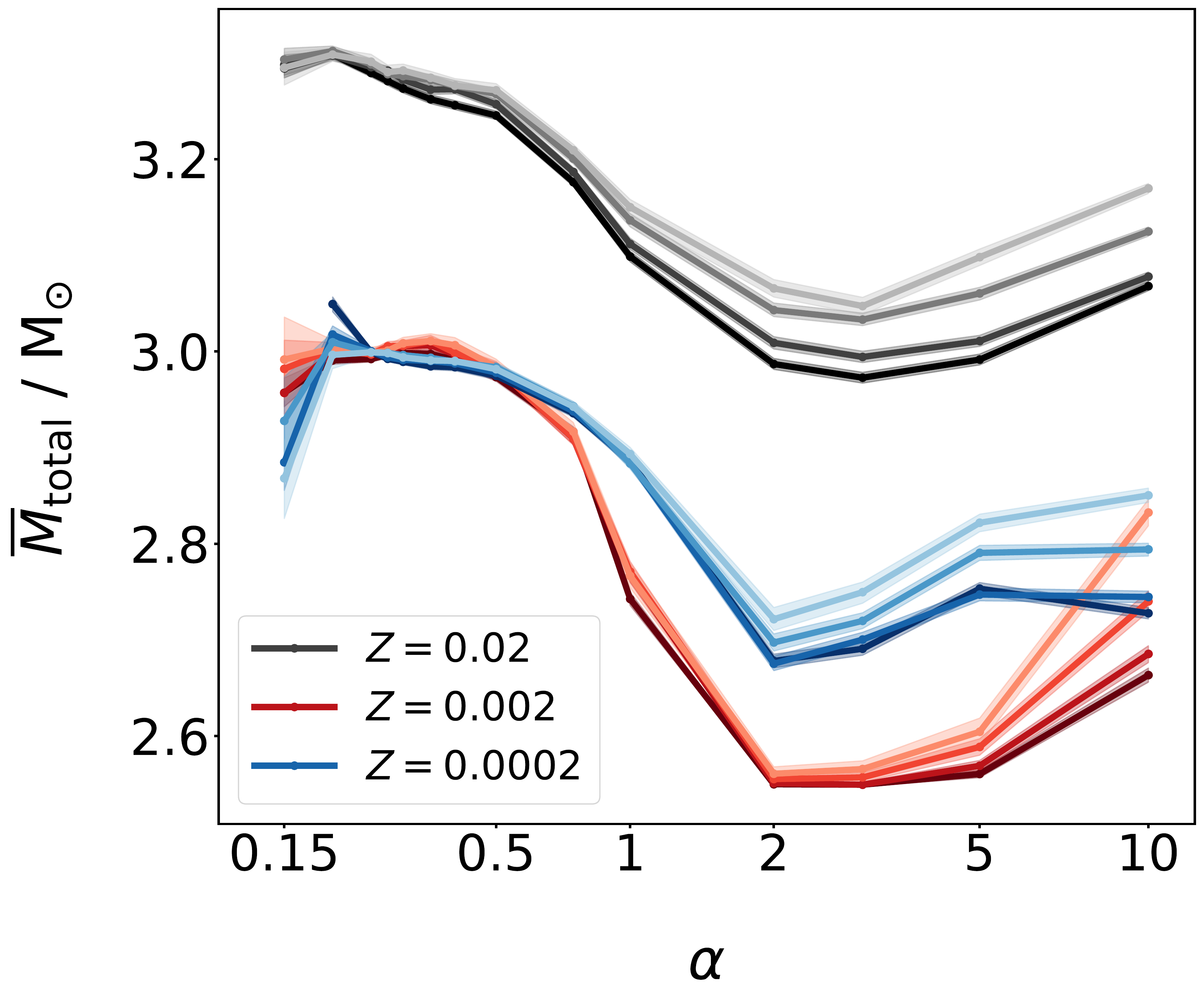}
    \caption{Mean total DNS mass.}\label{fig:total_mass}
  \end{subfigure}
    \hspace{15ex}
  \centering
    \begin{subfigure}{0.31\linewidth}
    \centering
    \includegraphics[width=\linewidth]{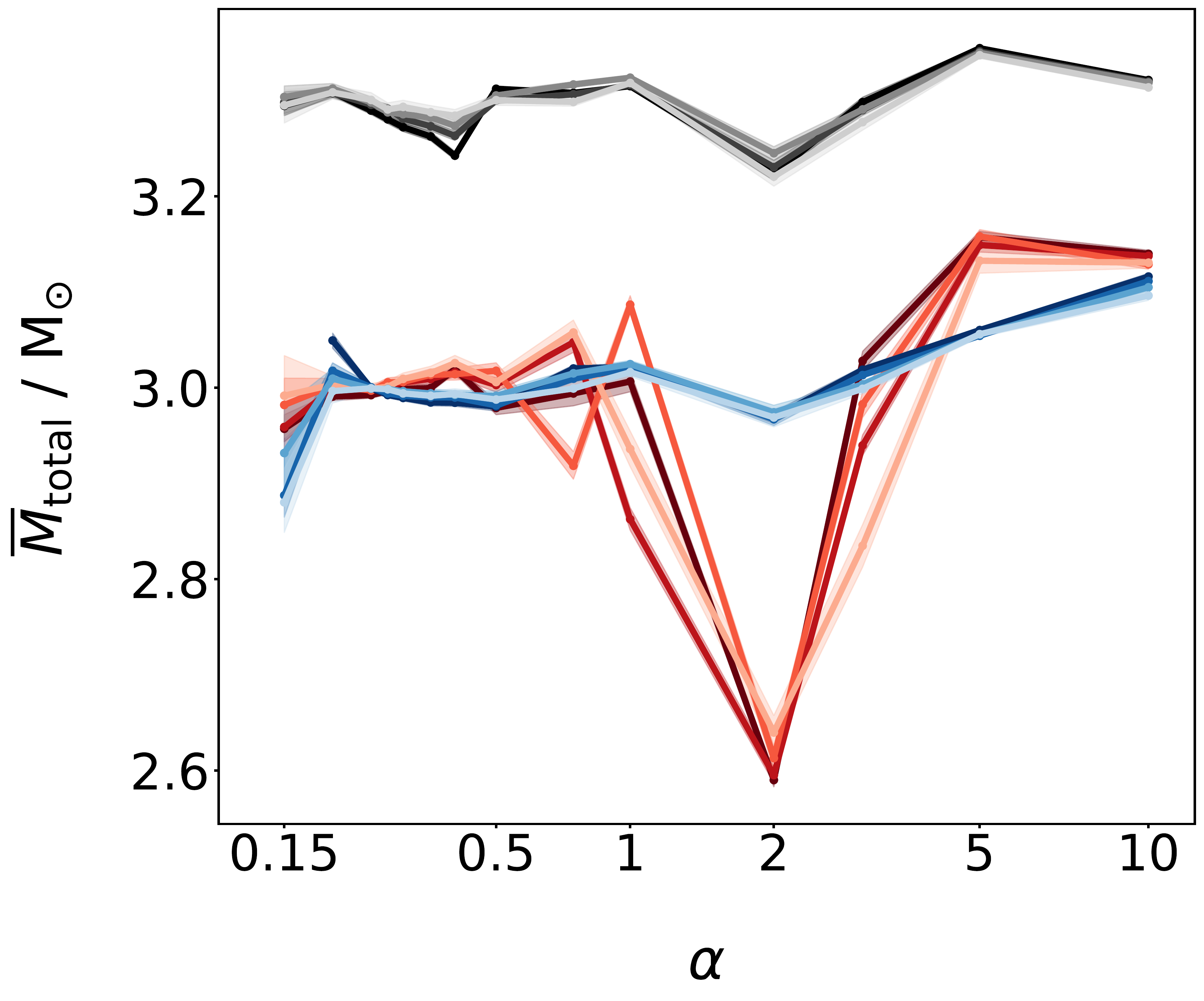}
    \caption{Mean total merging DNS mass.}\label{fig:total_mass_merge}
  \end{subfigure}
  \\  
  \caption{A comparison of the mean total DNS mass~$\overline{M}_\mathrm{total}$ (with bootstrapped $95$ per cent confidence intervals) as functions of~$\alpha$ for different ECSN kick dispersions, $\sigma_\mathrm{ECSN} \in \{0$, $7$, $15$, $26\} \, \textnormal{km}\,{\textnormal{s}}^{-1}$, at metallicities of $Z \in$ \{$0.02$, $0.002$, $0.0002$\}. In Panel~\subref{fig:total_mass_merge} DNSs are considered to merge if their delay times are less than~$13.8 \, \mathrm{Gyr}$. In the plots the colours are sequential with brighter shades representing higher ECSN kick dispersions and different hues representing different metallicities.}
  \label{fig:mass_stats}
\end{figure*}

The total number of DNSs decreases monotonically as the ECSN kick dispersion increases (Fig.~\ref{fig:all_m02}--\subref{fig:all_m0002}) because the probability of faster kicks increases, and these are more likely to lead to binary disruption. The number of merging DNSs does not decrease monotonically as the ECSN kick dispersion increases at all~$\alpha$ (Fig.~\ref{fig:merge_m02}--\subref{fig:merge_m0002}). For~$\alpha \leq 0.5$ the number of merging DNSs decreases monotonically with $\sigma_\mathrm{ECSN}$ and the local maximum decreases by a factor of about two to three from $\sigma_\mathrm{ECSN}=0$ to~$26\,\textnormal{km}\, {\textnormal{s}}^{-1}$. For~$Z = 0.002$ it is not possible to assess the effect of~$\sigma_\mathrm{ECSN}$ for~$0.75 \la \alpha \la 3.0$ because the number of merging systems in our simulations is too small, indicating that we do not resolve this regime properly.

\begin{table*} 
\caption{The maximum and minimum mean total DNS mass~$\overline{M}_\mathrm{total}$, and the associated model parameters, for all DNSs and for the subset of merging DNSs.} 
\label{tab:mass}
    \begin{tabular}{lcccccc}
\toprule
 All & \multicolumn{3}{c}{Minimum} & \multicolumn{3}{c}{Maximum} \\
\cmidrule(lr){2-4}\cmidrule(lr){5-7}
$Z$ & $\sigma_\mathrm{ECSN}\, / \, \textnormal{km}\, {\textnormal{s}}^{-1}$ & $\alpha$ & $\overline{M}_\mathrm{total} / \, \mathrm{M}_\odot$ & $\sigma_\mathrm{ECSN}\, / \, \textnormal{km}\, {\textnormal{s}}^{-1}$ & $\alpha$ & $\overline{M}_\mathrm{total} / \, \mathrm{M}_\odot$\\
\specialrule{.4pt}{2pt}{5pt}
$0.02$ & $0$ & $3$ & $2.973_{-0.006}^{+0.006}$ & $15$ & $0.20$ & $3.313_{-0.006}^{+0.005}$\\[4pt]
$0.002$ & $0$ & $3$ & $2.5494_{-0.0020}^{+0.0022}$ & $26$ & $0.35$ & $3.011_{-0.007}^{+0.008}$\\[4pt]
$0.0002$ & $7$ & $2$ & $2.675_{-0.008}^{+0.007}$ & $0$ & $0.20$ & $3.049_{-0.008}^{+0.007}$\\ [3pt]
\bottomrule
\toprule
Merging & \multicolumn{3}{c}{Minimum} & \multicolumn{3}{c}{Maximum} \\
\cmidrule(lr){2-4}\cmidrule(lr){5-7}
$Z$ & $\sigma_\mathrm{ECSN}\, / \, \textnormal{km}\, {\textnormal{s}}^{-1}$ & $\alpha$ & $\overline{M}_\mathrm{total} / \, \mathrm{M}_\odot$ & $\sigma_\mathrm{ECSN}\, / \, \textnormal{km}\, {\textnormal{s}}^{-1}$ & $\alpha$ & $\overline{M}_\mathrm{total} / \, \mathrm{M}_\odot$\\
\specialrule{.4pt}{2pt}{5pt}
$0.02$ & $26$ & $2$ & $3.221_{-0.010}^{+0.010}$ & $0$ & $5$ & $3.3548_{-0.0024}^{+0.0023}$\\[4pt]
$0.002$ & $0$ & $2$ & $2.590_{-0.007}^{+0.007}$ & $15$ & $5$ & $3.158_{-0.007}^{+0.007}$\\[4pt]
$0.0002$ & $26$ & $0.15$ & $2.88_{-0.03}^{+0.04}$ & $0$ & $10$ & $3.116_{-0.003}^{+0.003}$\\[3pt]
\bottomrule
    \end{tabular}
\end{table*}

In Fig.~\ref{fig:m_and_lambda} we show the effects of metallicity and our chosen prescription for~$\lambda$. The total numbers of DNSs and merging DNSs are lowest at a metallicity of~$0.002$ (Fig.~\ref{fig:all_ecap26} and ~\ref{fig:merge_ecap26}), in agreement with \citet{giacobbo_ecsne_dns}. Fig.~\ref{fig:form_lambda_002} and~\ref{fig:form_lambda_0002} show the total number of DNSs formed and Fig.~\ref{fig:merge_lambda_002} and~\ref{fig:merge_lambda_0002} display the number of DNSs that merge within the age of the Universe for constant~$\lambda = 0.5$ and for the $\lambda$-model of \citet{Claeysetal2014}. In Fig.~\ref{fig:merge_lambda_002} and~\ref{fig:merge_lambda_0002} both $\lambda$-prescriptions show local maxima in the number of merging DNSs. The local maximum for the number of merging NSs shifts to~$\alpha \approx 1$ when~$\lambda = 0.5$ but the general shape is the same as for the $\lambda$-prescription of \citet{Claeysetal2014}.

\subsection{Masses of DNSs}
\label{subsec:masses_of_DNSs}

\begin{figure*}
  \centering
    \begin{subfigure}{.9\linewidth}
    \centering
    \includegraphics[width=\linewidth]{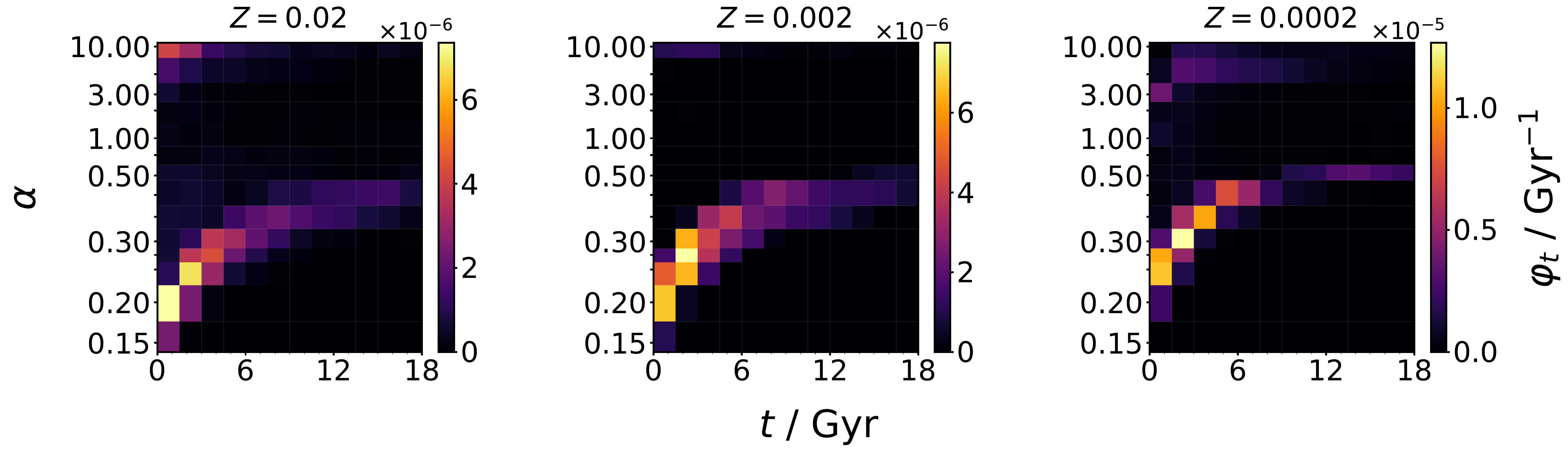}
    \caption{$\sigma_\mathrm{ECSN} = 0 \, \textnormal{km}\, {\textnormal{s}}^{-1}$.}\label{fig:delay_time_e0}
  \end{subfigure} \\
  \vspace{1ex}
    \begin{subfigure}{.9\linewidth} 
    \centering
    \includegraphics[width=\linewidth]{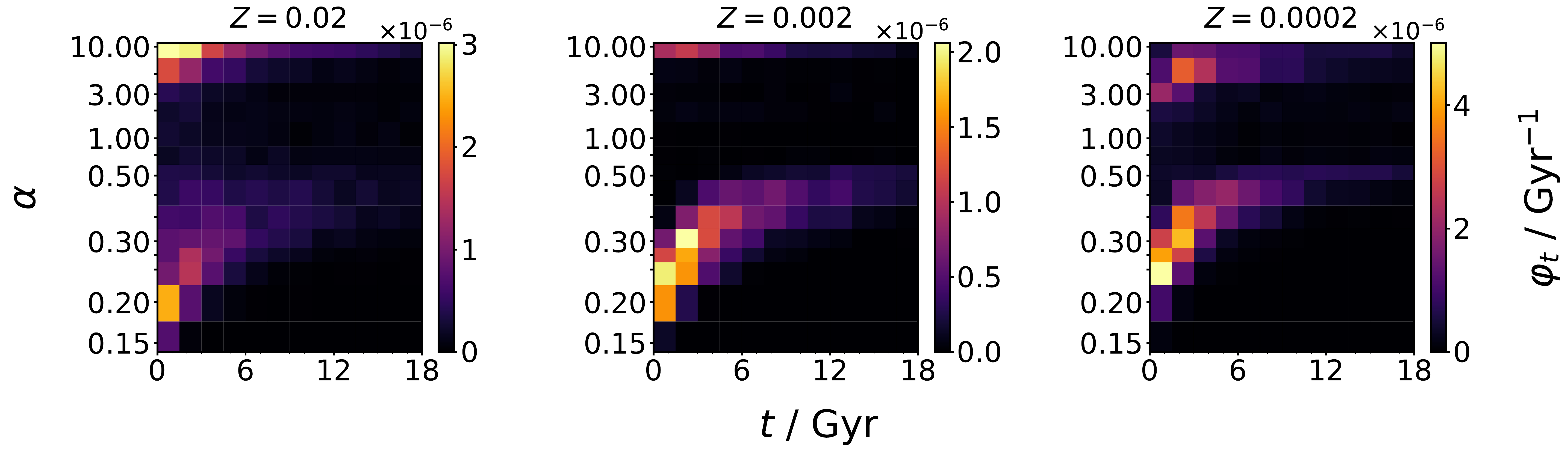}
    \caption{$\sigma_\mathrm{ECSN} = 26 \, \textnormal{km}\, {\textnormal{s}}^{-1}$.}\label{fig:delay_time_e26}
  \end{subfigure}\\
  \caption{Stacked histograms, for each~$\alpha$, of the DNS delay time~$t$ for delay times less than~$18 \, \textnormal{Gyr}$, where $\varphi_t$ is the probability density.}
  \label{fig:DelayTime}
\end{figure*} 

We bootstrapped the DNS mass distributions to generate confidence intervals for the sample means of the distributions, which reflect the non-normal nature of the distributions. Fig.~\ref{fig:mass_stats} shows the mean total DNS mass~$\overline{M}_\mathrm{total}$, with percentile bootstrapped $95$~per cent confidence intervals, as functions of~$\alpha$ for different models with ECSN kick dispersions $\sigma_\mathrm{ECSN} \in \{0$, $7$, $15$, $26\} \, \textnormal{km}\,{\textnormal{s}}^{-1}$ at metallicities~$Z \in \{0.02$, $0.002$, $0.0002$\}. The confidence intervals are calculated from $2\,000$ bootstrap re-samples of the simulated mass distributions of each grid. Each bootstrap re-sample is generated by randomly re-sampling with replacement the original simulated sample of DNS masses. The probability of re-sampling a mass is proportional to the probability of the initial binary parameters (Eq.~\ref{eqn:pdf_MqP}). Then the mean of each re-sample is calculated, resulting in $2\,000$ bootstrap estimates of the mean from which confidence intervals are deduced. The mean mass is not plotted for models that have extremely low total probability of DNS formation. The mean total DNS mass as function of~$\alpha$ (Fig.~\ref{fig:total_mass}) has a local maximum at a similar~$\alpha$ to the local maximum in the number of merging DNSs (Fig.~\ref{fig:merge_m02}--\subref{fig:merge_m0002}) and, similarly, also an increase at higher~$\alpha$. The total mass of the merging DNSs is generally higher than that of all DNSs. These findings are in line with the time it takes a DNS to merge being inversely proportional to its total mass (Eq.~\ref{eqn:numerical_fit_method}), with higher total DNS mass increasing the fraction of DNSs that merge, and the merging DNSs being further biased towards higher masses.
Table~\ref{tab:mass} shows a summary of the mean masses, with the mean merging total mass always being greater than (for~$\alpha > 2.0$) or equal to the mean total mass. This is very relevant in the context of GW$190425$, a merged DNS detected by gravitational waves. It has a mass significantly higher than that of known DNS systems \citep{GW190425}, as we discuss further in Sect.~\ref{subsec:discuss_ns_mass}.

The total DNS mass has a further dependency on the ECSN kick dispersion for~$\alpha \ga 1$, with faster kicks leading to higher total masses as lower-mass systems are more easily disrupted. The total DNS mass at~$\alpha \la 1$ and the total mass of merging DNSs do not strongly depend on the ECSN kick dispersion. The local minimum in the mean mass of merging DNSs at~$\alpha=2$ for~$Z=0.002$ coincides with a local minimum of the total DNS mass for all metallicities. However, this might be an artefact of the very small number of merging DNSs in the range~$0.75 \la \alpha \la 3.0$ for~$Z=0.002$, which is not resolved well with our grid resolution.

\subsection{Delay time distributions}
\label{subsec:delay_time_distributions}

\begin{figure*}
  \centering
    \begin{subfigure}{0.31\linewidth}
    \centering
    \includegraphics[width=\linewidth]{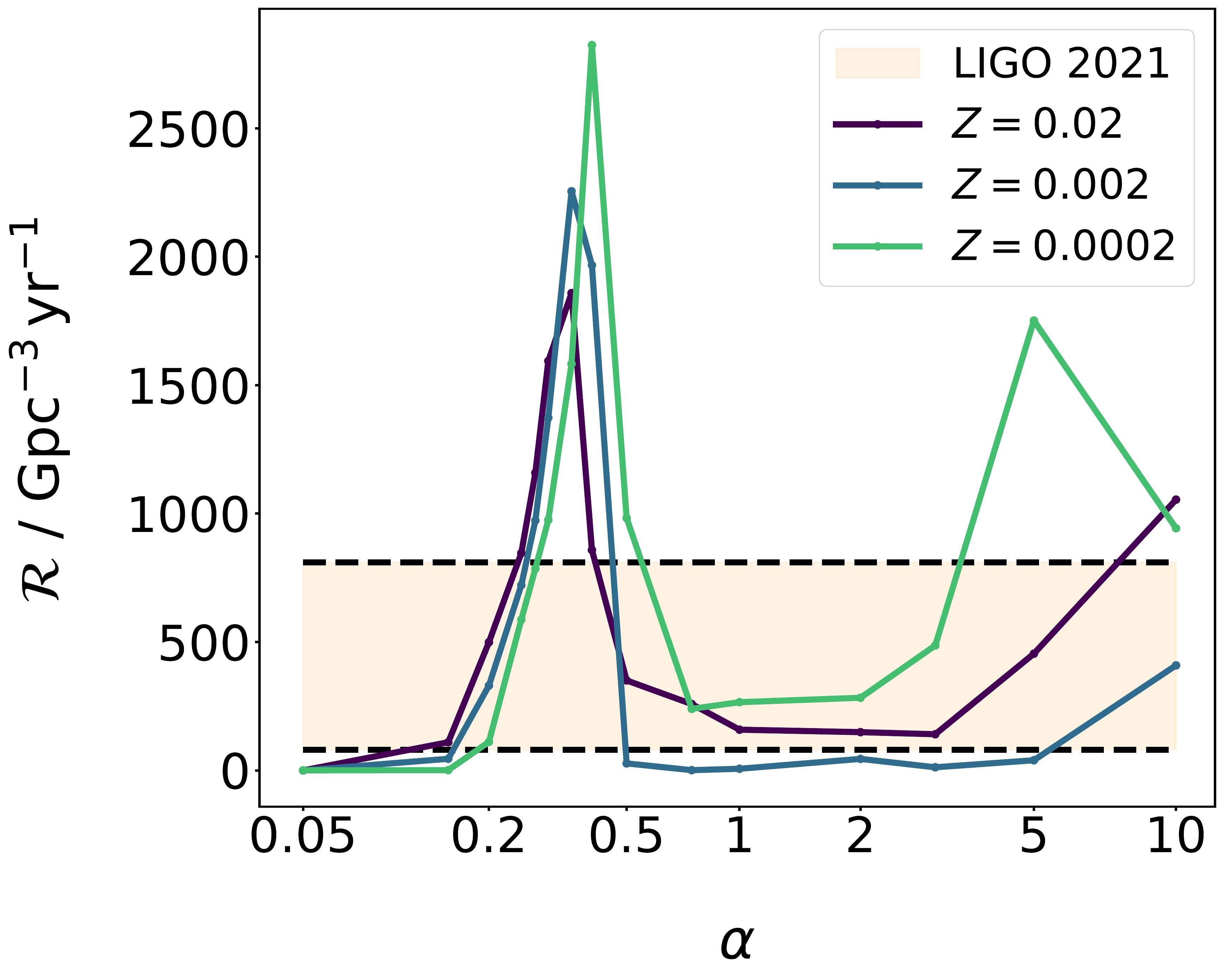}
    \caption{$\sigma_{\mathrm{ECSN}} = 0 \, \mathrm{km} \, \mathrm{s}^{-1}$.}\label{fig:r_ecsn_0}
  \end{subfigure}
    \hspace{2ex}
  \begin{subfigure}{0.31\linewidth}
    \centering
    \includegraphics[width=\linewidth]{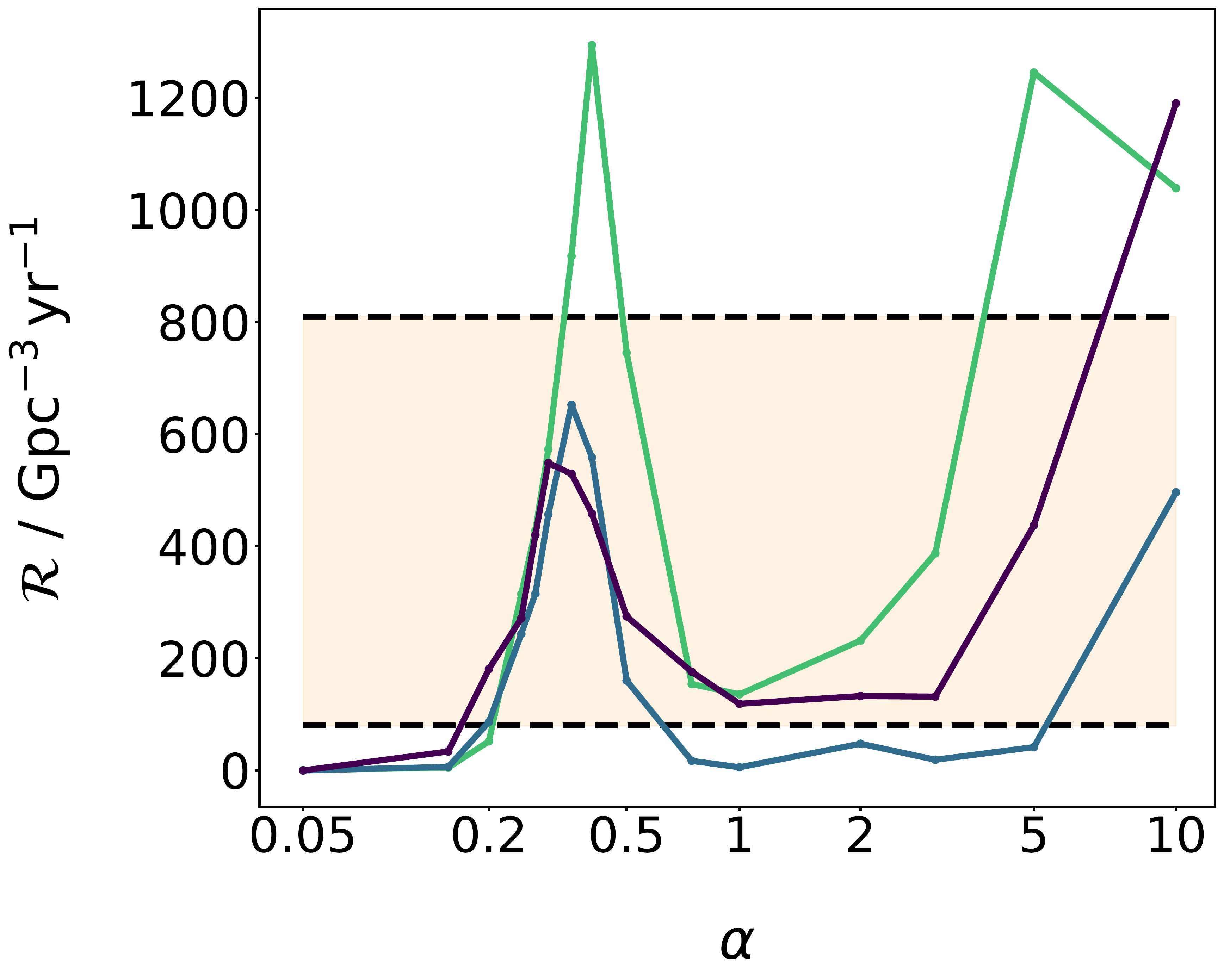}
    \caption{$\sigma_{\mathrm{ECSN}} = 26 \, \mathrm{km} \, \mathrm{s}^{-1}$.}\label{fig:r_ecsn_26}
  \end{subfigure}
      \hspace{2ex}
    \begin{subfigure}{0.31\linewidth}
    \centering
    \includegraphics[width=\linewidth]{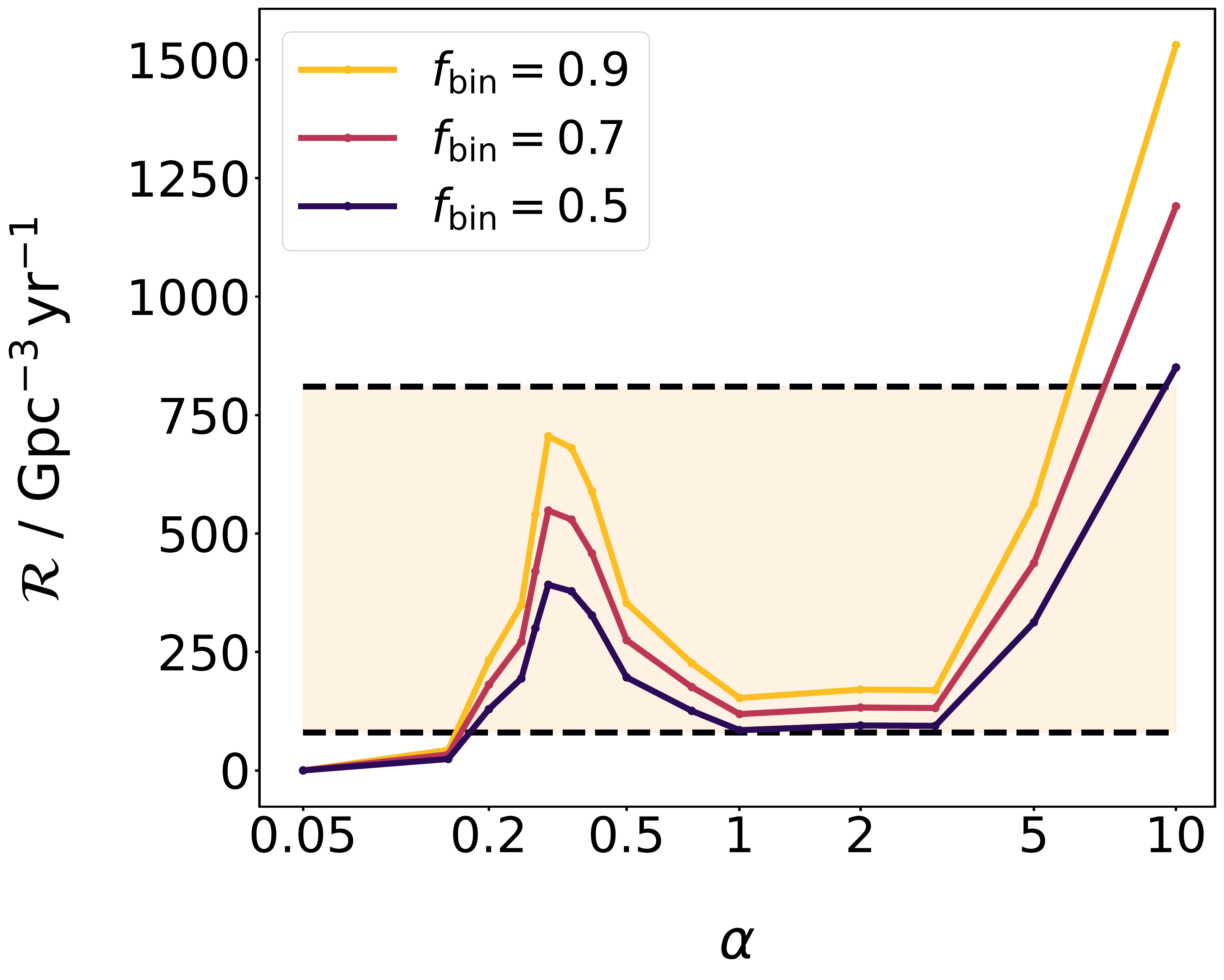}
    \caption{$\sigma_{\mathrm{ECSN}} = 26 \, \mathrm{km} \, \mathrm{s}^{-1}$, $Z = 0.02$.}\label{fig:r_fbin}
  \end{subfigure}\\
  \vspace{2ex}
  \centering
    \begin{subfigure}{0.31\linewidth}
    \centering
    \includegraphics[width=\linewidth]{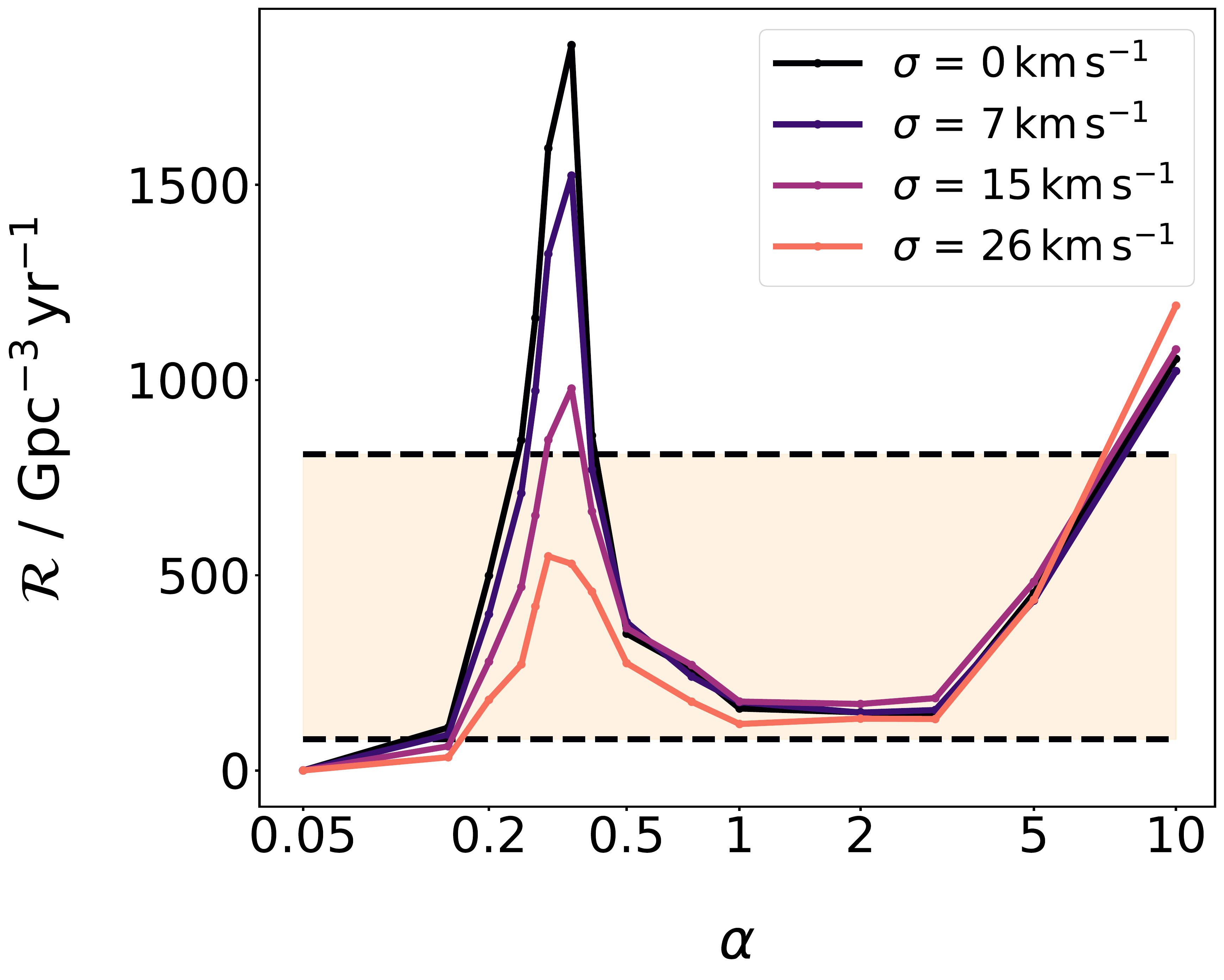}
    \caption{$Z = 0.02$.}\label{fig:r_z_02}
  \end{subfigure}
    \hspace{2ex}
  \begin{subfigure}{0.31\linewidth}
    \centering
    \includegraphics[width=\linewidth]{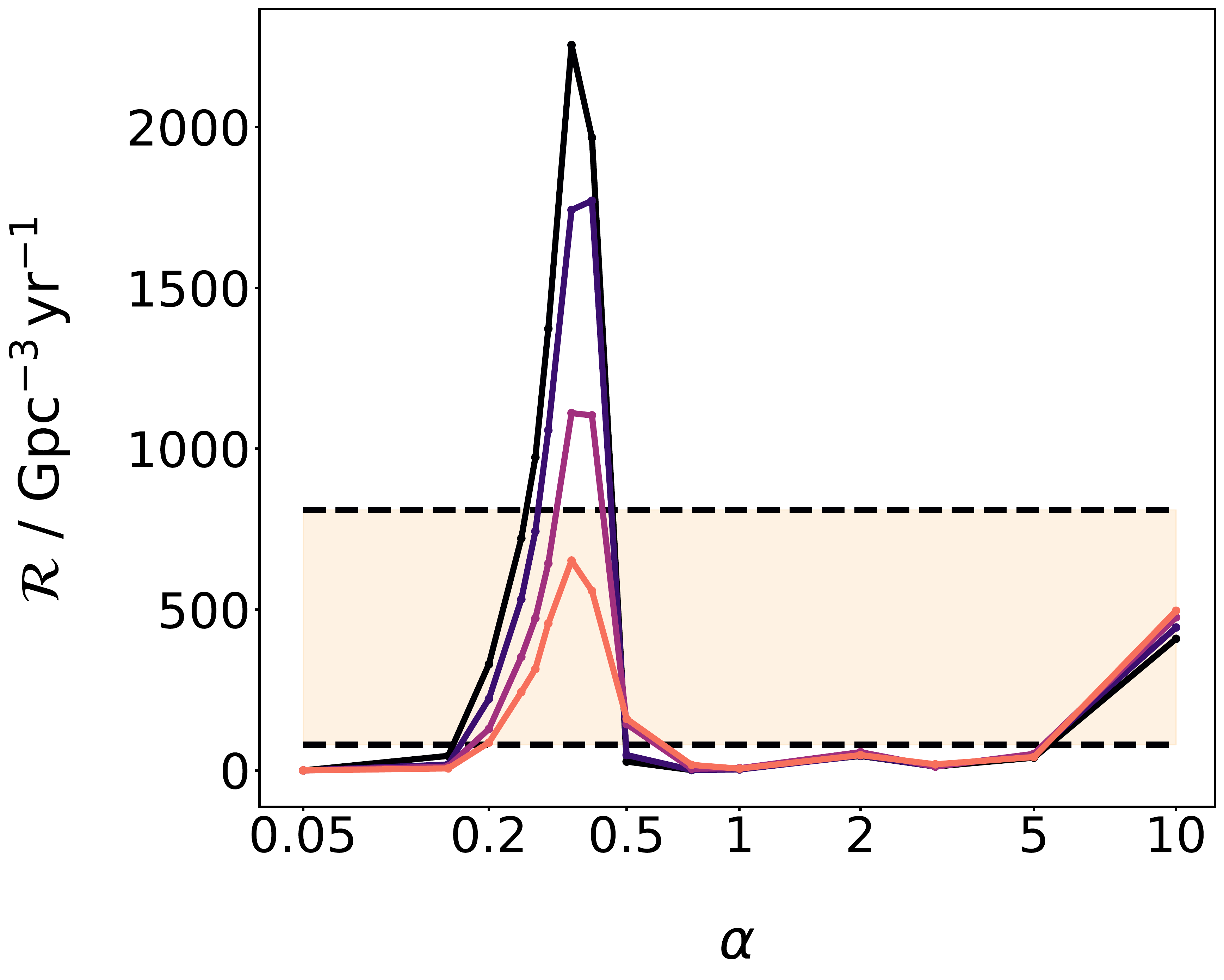}
    \caption{$Z = 0.002$.}\label{fig:r_z_002}
  \end{subfigure}
      \hspace{2ex}
    \begin{subfigure}{0.31\linewidth}
    \centering
    \includegraphics[width=\linewidth]{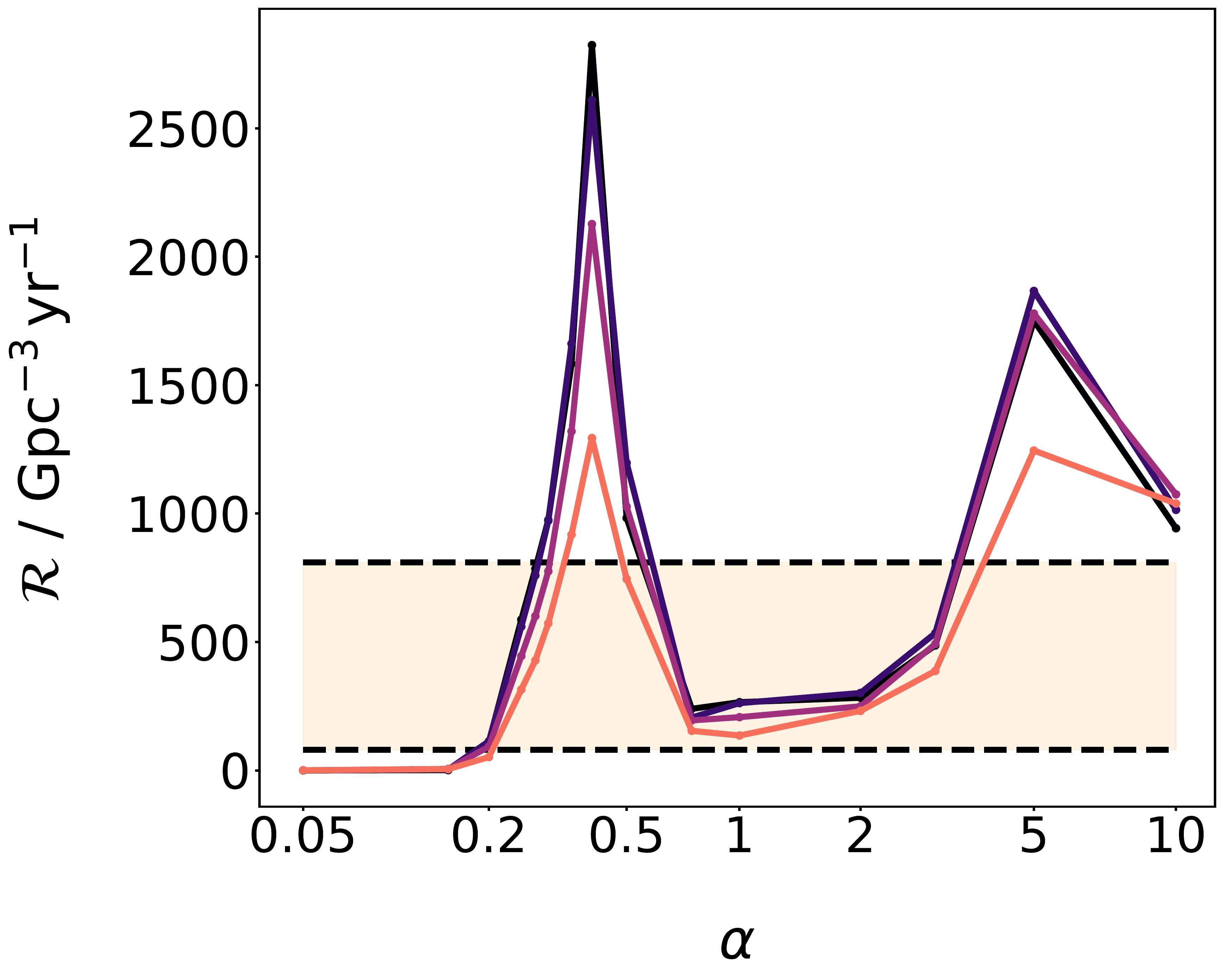}
    \caption{$Z = 0.0002$.}\label{fig:r_z_0002}
  \end{subfigure}\\
    \vspace{2ex}
    \begin{subfigure}{0.31\linewidth}
    \centering    
    \includegraphics[width=\linewidth]{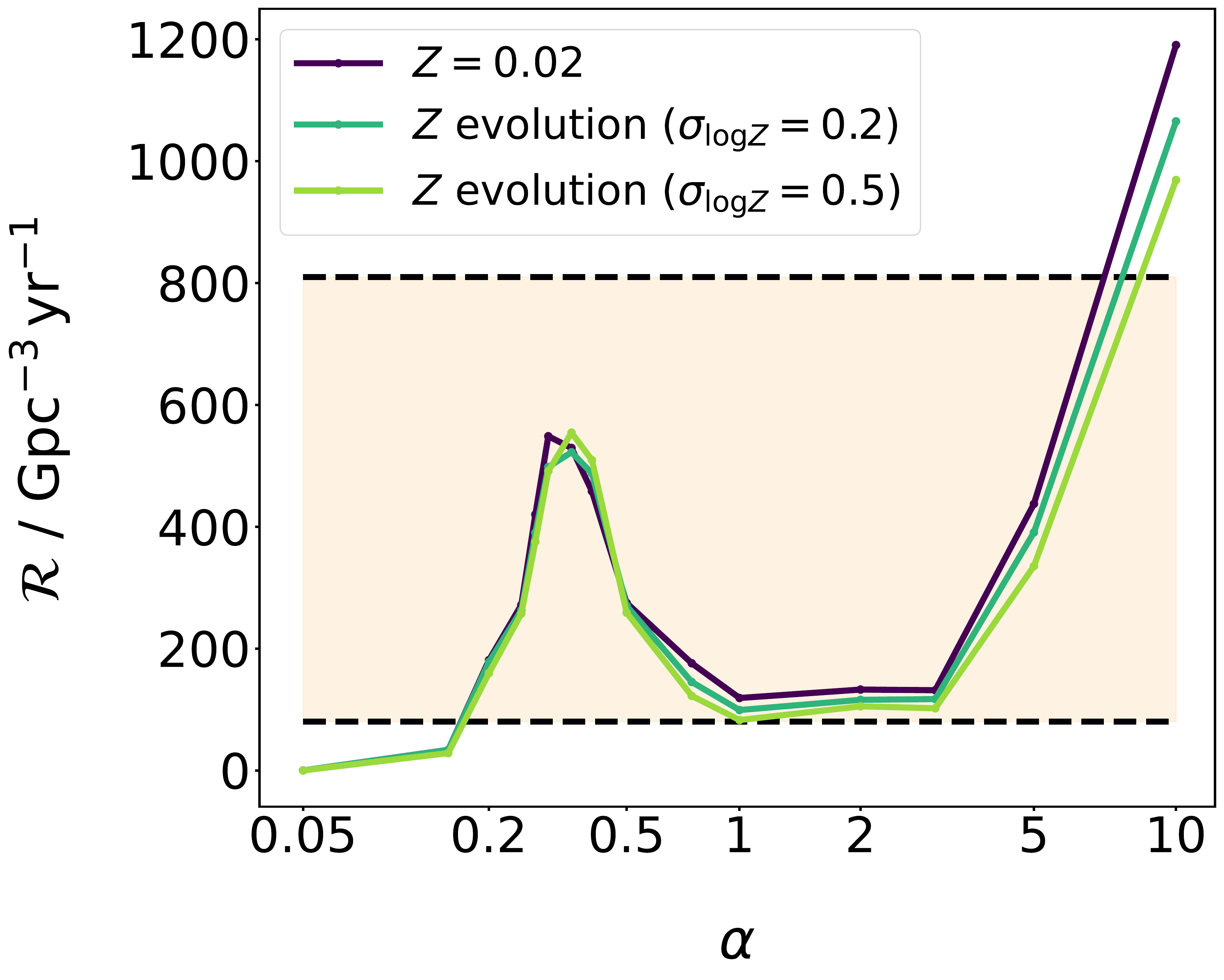}    \caption{$\sigma_{\mathrm{ECSN}} = 26 \, \mathrm{km} \, \mathrm{s}^{-1}$.}
    \label{fig:r_z_evol}
  \end{subfigure}\\    
  \caption{A comparison of the DNS merging rate~$\mathcal{R}$ as function of~$\alpha$ in different models with ECSN kick speed dispersions~${\sigma_\mathrm{ECSN} \in \{0, 7, 15, 26\} \, \textnormal{km}\,{\textnormal{s}}^{-1}}$ at metallicities of~$Z \in \{0.02$, $0.002$, $0.0002$\}. The DNS merging rate range of~$80$ to~$810 \, \mathrm{Gpc}^{-3}\,\mathrm{yr}^{-1}$ inferred by~\protect\cite{LIGO_BNS_merger_rate_2021} is also plotted. Panel~\subref{fig:r_z_evol} shows the rate calculated with~$\sigma_\mathrm{ECSN}= 26 \, \textnormal{km}\,{\textnormal{s}}^{-1}$ and the metallicity evolution model described in Sect.~\ref{subsec:merger_rate} with logarithmic metallicity dispersions~$\sigma_\mathrm{\log Z} \in$ \{$0.2$, $0.5\}$. We assume~$\protect f_\mathrm{bin} = 0.7$ except in Panel~\subref{fig:r_fbin}, which shows the simple scaling effect of varying~$f_\mathrm{bin}$.} 
  \label{fig:merger_rate}
\end{figure*}

We present the delay time distributions (DTDs) for $\sigma_\mathrm{ECSN}\,=\,0$ and~$\sigma_\mathrm{ECSN} = 26 \, \mathrm{km} \, \mathrm{s}^{-1}$ in Fig.~\ref{fig:DelayTime}. For most combinations of~$\alpha$ and~$\sigma_\mathrm{ECSN}$ the DTDs peak at short delay times and then decay to zero as the delay time increases. Between~$\alpha = 0.05$ and~$0.5$ the location of the peak shifts to longer delay times and the tail of the distribution lengthens. Between~$\alpha = 0.5$ and~$10$ the peak shifts back to short delay times. The general shape of the DTDs does not vary with metallicity, except for a slight shift of the peak towards longer delay times at~$\alpha \ga 3$ for~$Z=0.0002$ and an increase in the height of the peak at lower metallicities. 

In nearly all cases, a significant fraction of merging DNSs have delay times of~$t > 1\,\mathrm{Gyr}$. Therefore, we expect the detection of gravitational waves from merging DNSs that formed several $\mathrm{Gyr}$ ago, at epochs when star formation rates were higher than they are observed to be in the local Universe.

\subsection{Estimating the local DNS merging rate}
\label{subsec:estimating_the_local_DNS_merger_rate}

In Fig.~\ref{fig:merger_rate} we show the local DNS merging rate computed as described in Sect.~\ref{subsec:merger_rate}.
The local rate of merging DNSs as a function of~$\alpha$ has a local maximum at low~$\alpha$ followed by another increase at high~$\alpha$. Similarly to the unweighted number of merging DNS (Fig.~\ref{fig:merge_m02}--\subref{fig:merge_m0002}), the local merging rate monotonically decreases with increasing~$\sigma_\mathrm{ECSN}$ for~$\alpha \la 0.5$. The merging rate is not strongly affected by varying~$\sigma_\mathrm{ECSN}$ for~$\alpha \ga 1.0$. The local maxima in the local DNS merging rate for metallicities of~$0.02$,~$0.002$ and~$0.0002$ range from~$5.5 \times 10^{2}$ to~$1.9 \times 10^{3} \,\mathrm{Gpc}^{-3}\,\mathrm{yr}^{-1}$,~$6.5 \times 10^{2}$ to~$2.3 \times 10^{3} \,\mathrm{Gpc}^{-3}\,\mathrm{yr}^{-1}$ and~$1.3 \times 10^{3}$ to~$2.8 \times 10^{3} \,\mathrm{Gpc}^{-3}\,\mathrm{yr}^{-1}$, respectively. These local maxima occur at~$\alpha \approx 0.35$, with the maxima decreasing with increasing metallicity for almost any~$\sigma_\mathrm{ECSN}$. At all metallicities the minimum rate is at the lowest~$\alpha$ ($0.05$). The rate tends to zero as~$\alpha$ approaches zero because CE ejection is an essential part of the DNS formation channels we study.

Merging DNSs arise from binary systems that are formed with different metallicities. To account for this we assume a cosmic metallicity evolution with redshift as outlined in Sect.~\ref{subsec:merger_rate}. For $\sigma_\mathrm{ECSN} = 26 \,\textnormal{km}\,{\textnormal{s}}^{-1}$ we find a merging rate that is similar to the rate for~$Z = 0.02$ (Fig.~\ref{fig:r_z_evol}) because the linear evolution of the mean-logarithmic metallicity with redshift (Eq.~\ref{eqn:mean_Z}) leads to the mean metallicity spending the longest time in the coarse-grained region associated with our $Z = 0.02$~grids. Moreover, the SFRD (Eq.~\ref{eqn:sfrd}) has a local maximum at redshift~$z = 1.86$, which is likewise in the redshift region associated with our metallicity $Z = 0.02$~grids. The width of the logarithmic metallicity dispersion also affects the merging rate, with increasing~$\sigma_{\log Z}$ from~$0.2$ to~$0.5$ shifting the merging rate towards the $Z = 0.002$~rates (Fig.~\ref{fig:r_z_002} and~\ref{fig:r_z_evol}), because a larger dispersion leads to a wider metallicity distribution.

Overall, our computed local rates of merging DNSs are compatible with the DNS merging rate of~$80$ to~$810 \, \mathrm{Gpc}^{-3}\,\mathrm{yr}^{-1}$ inferred by~\protect\cite{LIGO_BNS_merger_rate_2021} over a large range of~$\alpha$, and more so at higher~$\sigma_\mathrm{ECSN}$ that produce lower local maxima. Near the local maximum, the computed rate can be higher than the upper limit derived from observations, as well as for~$\alpha > 5.0$. For~$\alpha < 0.15$ the computed rate is below the observational lower limit because most systems merge during CEE.

\section{DISCUSSION}
\label{sec:discussion}

We compare our findings with earlier work.

\subsection{Envelope ejection efficiency}
\label{subsec:discuss_alpha}

A sensitivity of the DNS merging rate to the assumptions for CEE phases has been shown in previous studies \citep[e.g.][]{comparison_chruslinka_2018,Santoliquido_september_2020,Broekgaarden2022}. Our study focuses on the region of $0.05\le \alpha \le 1$ with a higher resolution in~$\alpha$ compared to that of previous studies. For example, \cite{Santoliquido_september_2020} only checked~$\alpha=0.5$ and~$\alpha=1$ in this region, focusing more on~$\alpha > 1$ (motivated by \citealt{MapelliGiacobbo2018}, \citealt{CEE5} and \citealt{GiacobboMapelli2020}).

\cite{Santoliquido_september_2020} found a local maximum in the rate of merging DNSs at~$\alpha = 7$, though their investigation was limited to only one assumed natal kick distribution, while with other parameter variations they focused only on~$\alpha=1$ and~$\alpha=5$. We find a second peak, at~$\alpha=5$, but only at low metallicity ($Z = 0.0002$). It is possible that a similar peak is present at other metallicities but we miss it because of our sparser sampling at high~$\alpha$ than near the first peak.

Additional details that might affect the envelope ejection are the role of recombination energy \citep*{CEErecombination2015,CEErecombination2018} and thermal energy \citep{comparison_kruckow_2018}, as well as the envelope structure \citep{lambda_dewi_2000,lambda_tauris_2001}. A metallicity-dependent $\lambda$-prescription \citep[e.g. ][]{binding_energy_xu_li,binding_energy_loveridge,binding_energy_klencki}, for example, might affect our results. Our simulations with~$\lambda=0.5$ (Fig.~\ref{fig:m_and_lambda}) demonstrate that the dependence of the number of merging DNSs on~$\alpha$ remains qualitatively unchanged but at a specific~$\alpha$ there can be a large dependence on the envelope structure parameter, reinforcing the crucial dependence on the uncertain CE ejection physics (encapsulated here in~$\alpha$). 

\cite{Santoliquido_september_2020} noted that further exploration of the $\alpha$--$\sigma$~space would be desirable in the future and our study is just such an exploration. Our results agree with the orders-of-magnitude sensitivity to~$\alpha$ that \cite{Santoliquido_september_2020} found for the DNS merging rate. Different studies aim to estimate~$\alpha$ from detailed hydrodynamic simulations. Results vary between~$\alpha \approx 5$ \citep{CEE5} and~$\alpha \approx 0.1$--$2.7$ \citep{CEE6}. A recent study of white dwarf binaries by \cite{alphaScherbakFuller} suggests~$0.2 \la \alpha \la 0.4$. In conclusion, while progress has been made, the uncertainty regarding~$\alpha$ remains high.

\subsection{Natal kicks}
\label{subsec:discuss_kicks}

The peak in the DNS merging rate as a function of~$\alpha$ is sensitive to the assumed kick speed distribution. The peak becomes insignificant at a high enough dispersion ($\sigma_\mathrm{ECSN} \geq 265\,\mathrm{km}\,\mathrm{s}^{-1}$). The rate at~$\alpha>1$ also depends on~$\sigma_\mathrm{ECSN}$ because of merging DNSs that include an ECSN in their formation history.

\cite{ecsne_low_kick} proposed that ECSNe and possibly core-collapse supernovae (CCSNe), with low-mass iron cores, produce NSs with slow natal kicks. Our results do not specifically require ECSNe, but only that some CCSNe occur with slow kicks. \cite{KickECSNeHD2018} find very slow kicks in hydrodynamic simulations of ECNSe but propose that CCSNe, with low-mass iron cores, result in faster kicks because the density profiles near their cores are shallower than in oxygen-neon cores of ECSN progenitors.

\cite{VignaGomez2018} and \cite{GiacobboMapelli2020} have suggested a revised natal kick prescription that gives slow kicks for ECSNe as well as for stars stripped of their envelopes in binaries. These include Type~Ib/c CCSNe, for which we assumed~$\sigma = 265\,\mathrm{km}\,\mathrm{s}^{-1}$, similarly to Type~II CCSNe. Ultra-stripped SNe (USSNe), which are Type~Ib/c SNe with ejecta masses  much lower than the remnants they form, might have low kicks as well \citep*{review_2.3_tauris}. Slower kicks in Type~Ib/c CCSNe might be favorable for merging DNSs. We hypothesise that a peak in the merging rate of DNSs as a function of~$\alpha$ should appear in population synthesis simulations that include a substantial number of NSs formed in SNe with slow kicks. The prescription suggested by \cite{GiacobboMapelli2020} is implemented in a later version of \textsc{binary\_c} than the one we use here, and its effect on the rate of merging DNSs will be investigated in a future study, including low-velocity kicks in USSNe.

\subsection{NS mass distributions}
\label{subsec:discuss_ns_mass}

Observational studies \citep{bimodalnsmass2,bimodalnsmass, bimodalnsmass2018} of Galactic pulsars suggest that there is a bimodal NS mass distribution, potentially owing to different formation channels. On the other hand, the extragalactic mass distribution of NSs observed in gravitational waves is consistent with a uniform distribution \citep{mass_GW_inference_landry}, though this inference is based on a small number of DNS systems. The mass distributions of individual NSs in our simulated DNS populations are non-overlapping bimodal distributions when~$\alpha < 1.0$, with high- and low-mass components consistent with observational studies of pulsars (Appendix~\ref{sec:appendixc}).

Different masses of merging neutron stars were inferred in two gravitational wave events, GW$170817$ ($M_\mathrm{total}=2.74_{-0.01}^{+0.04}\,\mathrm{M}_\odot$; \citealt{GW170817_discovery}) and GW$190425$ ($M_\mathrm{total}=3.4_{-0.1}^{+0.3}\,\mathrm{M}_\odot$; \citealt{GW190425}). The inferred total mass observed in GW$170817$ is consistent with the mass distribution of Galactic DNSs (\citealt*{GalacticDNSs2019}; \citealt{GW190425}), while the total mass of GW$190425$ is significantly higher. We propose a tentative explanation: the mass distribution of merging DNSs differs from the overall DNS mass distribution. This is possible if the merging DNSs are only a relatively small fraction of all DNSs and this is the case in our simulations with~$\alpha \ga 0.4$. When the fraction of DNSs that merge within the age of the Universe is close to unity, as we find when~$\alpha \ll 1$, there is a difference between the mass inferred for GW$190425$ and the Galactic DNS masses. We find that the mean merging DNS masses deviate from the mean total DNS masses when~$\alpha \ga 1$ (Fig.~\ref{fig:mass_stats}).

Another possibility for the discrepancy between the masses of Galactic pulsars compared to extragalactic merging DNSs is that they are derived from disparate observations, which have different selection effects and biases. Galactic pulsars are measured in radio surveys, which are impacted by several effects like the beaming fraction \citep{review_selection_effects_tauris}, and biased towards systems which contain a radio-alive NS\footnote{In Appendix~\ref{sec:appendixd} we also compare the observed period-eccentricity distribution of Galactic pulsars to our simulated distributions.}. Gravitational wave measurements, however, are biased to more local systems and towards high masses. Future pulsar surveys and space-based gravitational wave detectors might enable the joint observation of the same DNS population in both radio and gravitational waves \citep{review_selection_effects_chattopadhyay}.

\subsection{Metallicity effects}
\label{subsec:discuss_metallicity}

Our continuum of metallicities is generated by a rather crude weighting because we only simulate three metallicities. For the purpose of stressing the sensitivity of the DNS merging rate to the CE ejection parameter~$\alpha$, this is sufficient. Further refinement of our results would require additional metallicities, especially in the regime of~$Z > 0.002$ because higher metallicities have a larger influence on the local observed rate when assuming linear mean logarithmic metallicity evolution. The contribution of different metallicities might also depend on~$\alpha$ in a non-trivial manner \citep{Santoliquido_september_2020}. Another possible metallicity effect is that the IMF depends on redshift \citep{varying_IMF_Chruslinska_2020}, while we assume a constant IMF. 

\cite{Laplaceetal2020} find, at low metallicity, that post-RLOF (Roche-lobe overflow) stars expand to radii much larger than those computed by \textsc{binary\_c} and similar codes (that use the prescriptions given by \citealt*{Hurley2000}) and claim that the interactions leading to DNS formation would be altered. A further implication, suggested by~\cite{Laplaceetal2020}, is a stronger metallicity dependence for DNS formation compared to other studies \citep{metallicity_effect_on_mergers_Neijssel_2020}.

The post-RLOF expansion to large radii discussed by~\cite{Laplaceetal2020} can be a result of a very tenuous and loosely-bound low-mass hydrogen envelope, which can be ejected by a small amount of energy, such that the further interaction does not cause a significant decrease in the orbital separation. Furthermore, this is not necessarily an effect of low metallicity, because post-RLOF wind mass loss might be overestimated at high metallicity \citep{HeliumStarWinds2017,HeliumStarWinds2019}. Nonetheless, the impact of post-RLOF expansion on binary interactions and rates of merging of compact objects should be investigated in detail with updated models for post-RLOF stripped stars.

\section{CONCLUSIONS}
\label{sec:conclusions}

We find that the local rate of merging DNSs depends critically on the efficiency of transferring orbital energy to the ejection of the common envelope (Fig.~\ref{fig:all_m},~\ref{fig:m_and_lambda} and~\ref{fig:merger_rate}). With an efficiency factor~$0.3 \, \la \, \alpha \, \la \, 0.4$ we find that the rate is highest and, in our understanding, this is the result of two competing effects.
\begin{enumerate}
    \item A low envelope-ejection efficiency, $\alpha \la 0.4$, requires a large fraction of orbital energy to be deposited in the ejection process (high~$\alpha$) and so the final separations and the time to merge are shorter. The fraction of DNS systems that merge is very high, close to unity, at low~$\alpha$.
    \item A high envelope ejection efficiency ($\alpha\ga 0.4$) means that a larger fraction of systems that enter a CEE phase survive as binaries rather than merging fatally and being prevented from contributing to the merging DNS gravitational wave sources.
\end{enumerate}
This peak in the DNS merging rate depends on the natal kick speed during the ECSN explosion. The height of the peak decreases with increasing speed dispersion in the ECSN natal kick distribution from~$\sigma_\mathrm{ECSN} = 0$ to~$26\,\textnormal{km}\, {\textnormal{s}}^{-1}$. This result highlights the importance of understanding the physical source of NS natal kicks and whether various types of supernovae have different kick speed distributions.

The mass distributions of DNS systems also depend on the CE ejection efficiency. The mass distributions of individual NSs comprising DNSs are non-overlapping bimodal distributions at low~$\alpha$ before transitioning to overlapping bimodal distributions with increasing~$\alpha$ and finally becoming essentially unimodal at high~$\alpha$. The mean total mass of merging DNSs is larger than the mean total mass of all DNS systems for high $\alpha$.

Our results demonstrate the high uncertainty in estimates of the rate of merging of DNSs and especially the uncertainty that arises from CEE physics. This motivates the development of more sophisticated CE models than are currently used. Such CE models could more naturally accommodate additional energy sources and non-dynamical inspiral time-scales than the canonical $\alpha$-prescription. Our findings are relevant in the context of the increasing number of merging DNSs expected to be detected as gravitational wave sources in the near future.

\section*{Acknowledgments}

We thank an anonymous referee for a constructive review of the paper.
AMT thanks Robert~J. Eady, J\'ozef Niedbalec, Janina Niedbalec, Erik~A. Rosenberg and Marc van~der~Sluys for helpful discussions.
AG acknowledges support from a grant by the Pazy Foundation.
RGI thanks the Science and Technology Facilities Council (STFC) for funding his Rutherford fellowship under grant ST/L$003910/1$ and consolidated grant ST/R$000603/1$.
CAT thanks Churchill College for his fellowship.

\section*{Data Availability}

The simulation outputs necessary to reproduce our results are summarised in tables that are available at \href{https://zenodo.org/record/7811486#.ZDJt6I5BzJU}{https://zenodo.org/record/7811486\#.ZDJt6I5BzJU}.

\bibliographystyle{mnras}
\bibliography{CEEalpha} 

\appendix
\section{Merging time numerical fit}
\label{sec:appendixa}

\begin{figure*}
  \centering
    \begin{subfigure}{0.32\linewidth}
    \centering
    \includegraphics[width=\linewidth]{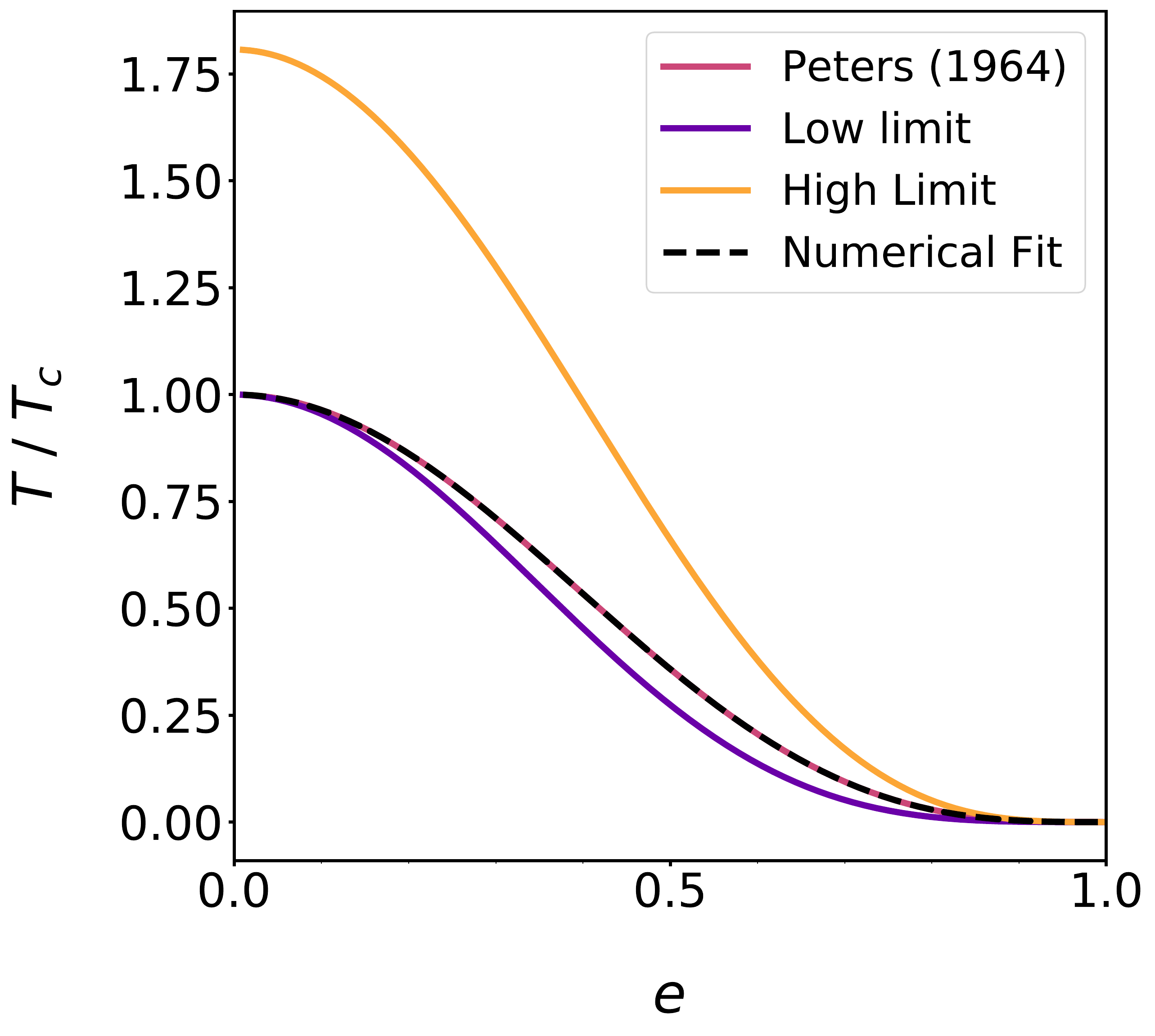}
    \caption{}\label{fig:peters_comparison_linear}
  \end{subfigure}
    \hspace{1ex}
  \begin{subfigure}{0.32\linewidth}
    \centering
    \includegraphics[width=\linewidth]{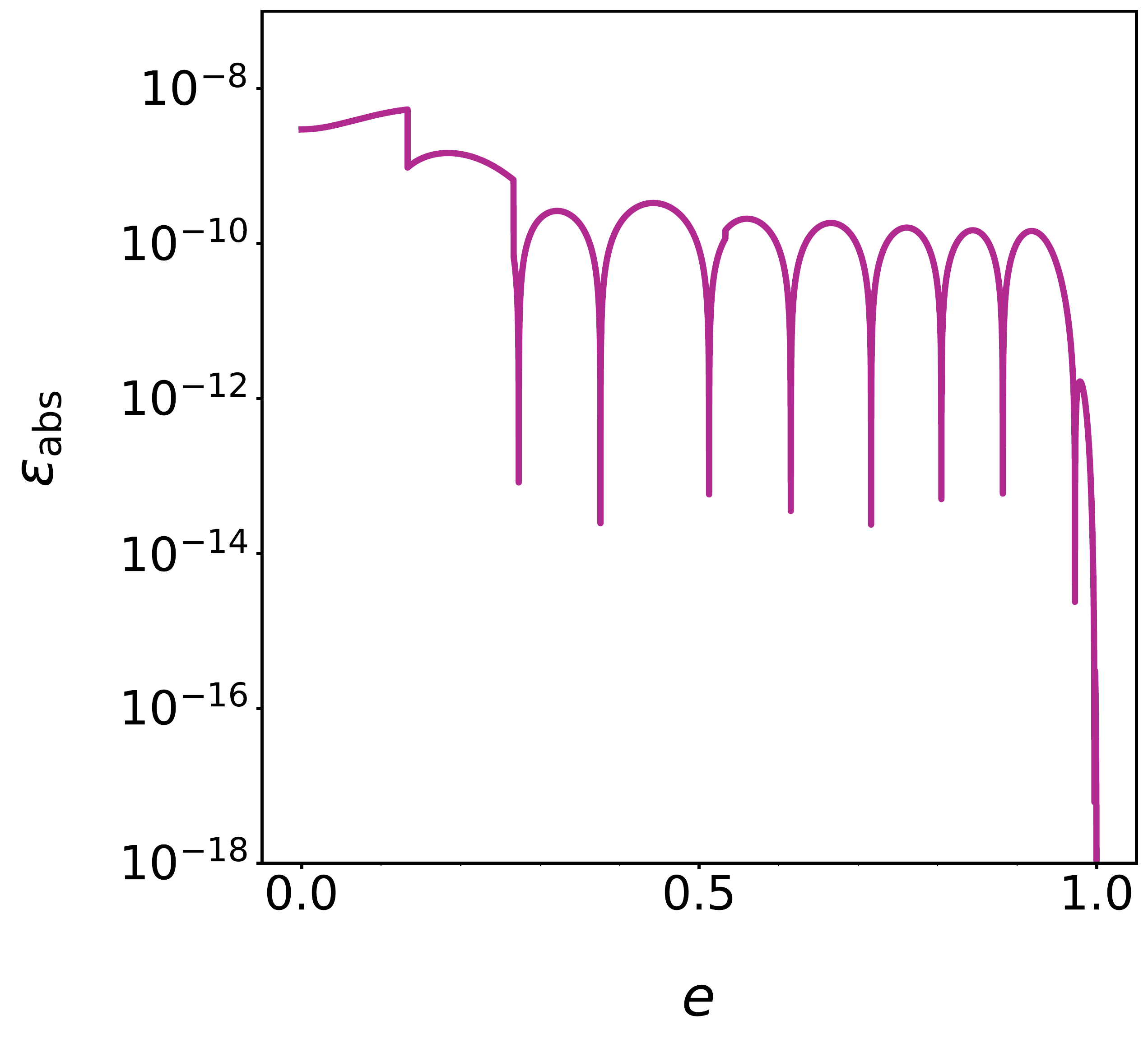}
    \caption{}\label{fig:peters_fit_errors}
  \end{subfigure}
      \hspace{1ex}
    \begin{subfigure}{0.32\linewidth}
    \centering
    \includegraphics[width=\linewidth]{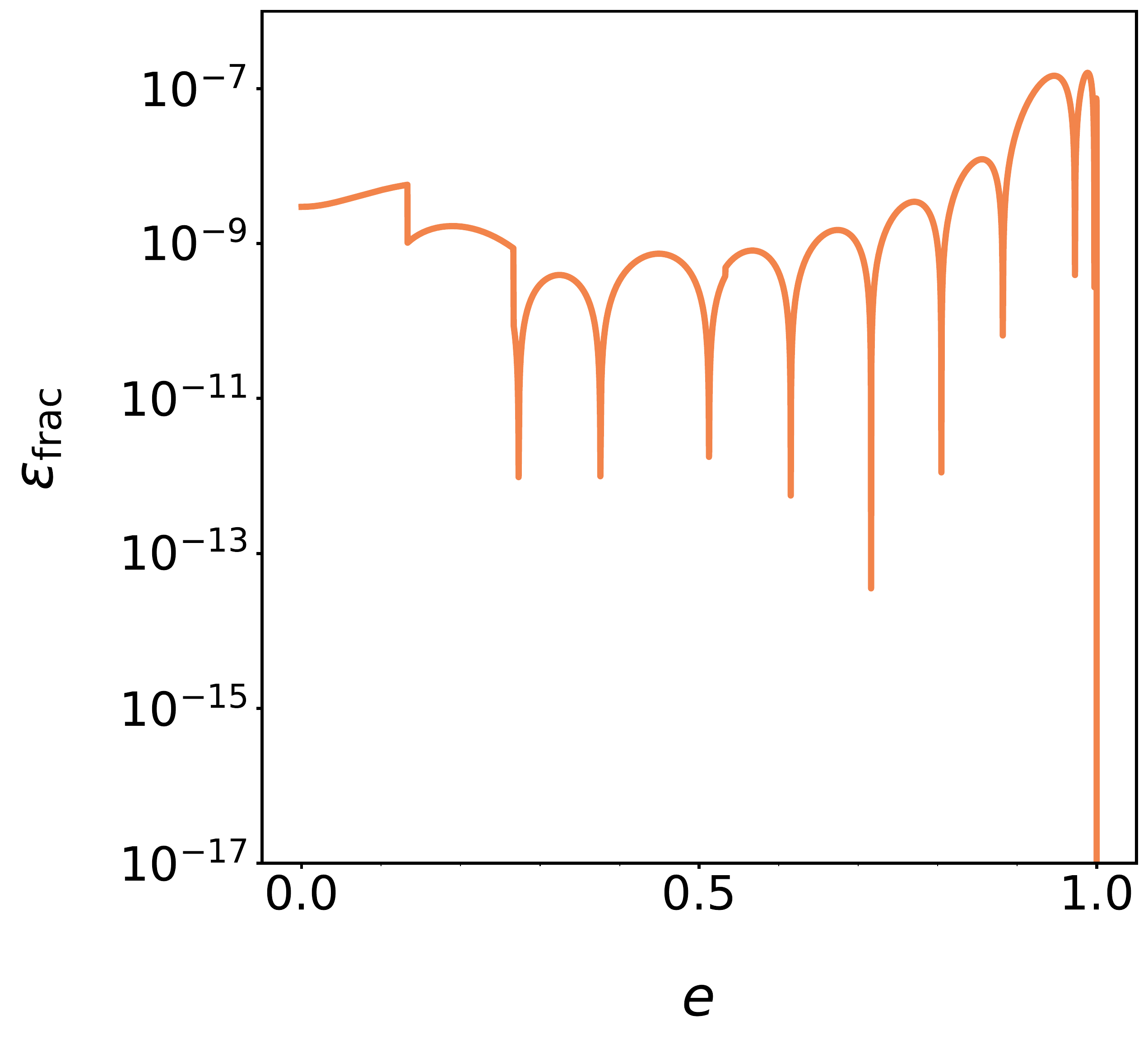}
    \caption{}\label{fig:peters_fit_frac_errors}
  \end{subfigure}\\
  \caption{A comparison of~\citeauthor{GW_rad}' (\citeyear{GW_rad}) merging time formula, its high and low limit approximations and our numerical fit. Panel~\subref{fig:peters_comparison_linear} is a linear plot of the ratio of the merging time to the circular orbit merging time~$T / T_\mathrm{c}$ as a function of eccentricity~$e$. Panels~\subref{fig:peters_fit_errors} and~\subref{fig:peters_fit_frac_errors} are plots of the absolute errors~$\varepsilon_\mathrm{abs}$ and fractional errors~$\varepsilon_\mathrm{frac}$ between~\citeauthor{GW_rad}' (\citeyear{GW_rad}) formula and our numerical fit as a function of eccentricity.}
  \label{fig:peters_comparison}
\end{figure*}

\begin{table*} 
\caption{The fitted constants~$\beta_{i}$, where $i \in \{1, 2,..., 10\}$, of our numerical fit (Eq.~\ref{eqn:numerical_fit}).} 
\label{tab:constants}
    \begin{tabular}{ccccc}
\toprule
$\beta_{1}$ & $\beta_{2}$ & $\beta_{3}$ & $\beta_{4}$ & $\beta_{5}$\\
\midrule
$-2.63021$ & $3.80120$ & $-4.45283647437033$ & $5.098282166215911$ & $-5.333378774161088$\\
\bottomrule\\ 
\toprule
$\beta_{6}$ & $\beta_{7}$ & $\beta_{8}$ & $\beta_{9}$ & $\beta_{10}$\\
\midrule
$5.035018671298129$ & $-3.874433738575514$ & $2.21156856755609$ & $-0.803477742235374$ & $0.1412084979057513$ \\
\bottomrule
    \end{tabular}
\end{table*}

The time to merge for two point masses orbiting each other can be approximated by applying the quadrupole formula for the energy loss of a system due to gravitational wave radiation to a Newtonian orbit as outlined by \citet{GW_rad_peters_1963} and \citet{GW_rad}. The time to merge for two point masses in a circular orbit is 
\begin{equation}
    T_\mathrm{c}(a_0) = \frac{5}{256} \frac{c^5}{G^3} \frac{a_{0}^{4}}{m_1 m_2 (m_1 + m_2)}\: , \label{eqn:circular_merge}
\end{equation}
where $T_\mathrm{c}$ is the time for a circular binary to merge after its formation,~$a_0$ is the initial semi-major axis of the binary orbit and~$m_1$ and~$m_2$ are the masses of the binary components. The time to merge~$T$ for two point masses in a more general elliptic orbit is 
\begin{multline}
    \frac{T(a_0,e_0)}{T_\mathrm{c}(a_0)} = \frac{48}{19} \frac{\big(1 - e_{0}^2 \big) ^ 4}{e_{0}^{48/19}\big[1 + (121/304)e_{0}^2 \big] ^{3480/2299} } \\
    \times \displaystyle\int\limits_{0}^{e_{0}} \mathrm{d}e  \frac{e^{29/19}\big[1 + (121/304)e^2 \big] ^{1181/2299} }{\big(1 - e^2 \big) ^{3/2}}\: ,\label{eqn:peters}
\end{multline}
where $e_0$ is the initial eccentricity of the orbit. The time to merge for a binary decreases as the eccentricity increases. Gravitational wave radiation also circularises binary orbits. Eq.~\eqref{eqn:peters} has low~$e_{0} \ll 1$ and high~$\epsilon \equiv |1-e_{0}| \ll 1$ eccentricity limits given by \citet{GW_rad} of
\begin{align}
    \frac{T(a_0,e_0)}{T_\mathrm{c}(a_0)} &= \frac{\big(1 - e_{0}^2 \big) ^ 4}{\big[1 + (121/304)e_{0}^2 \big] ^{3480/2299} }\: + \: \mathcal{O}\big( e_{0}^2 \big)\: , \label{eqn:low_limit}\\
    \frac{T(a_0,e_0)}{T_\mathrm{c}(a_0)} &= (768/425)\big(1 - e_{0}^2 \big) ^{7/2}+ \: \mathcal{O}\big( \epsilon^{4} \big)\: , \label{eqn:high_limit}
\end{align}
where we have determined the order of the residuals of these limiting functions. The first few terms of the series expansion of Eq.~\eqref{eqn:peters} about the points~$e_{0} = 0$ and~$e_{0} = 1$ are given by
\begin{align}
    \frac{T(a_0,e_0)}{T_\mathrm{c}(a_0)} &= 1 -\frac{157}{43} e_{0}^{2} + \frac{220521}{42656} e_{0}^{4} + \displaystyle\sum\limits_{i=3}^{\infty}\gamma_{i}e_{0}^{2i} \: , \label{eqn:low_limit_series}\\
   \frac{T(a_0,e_0)}{T_\mathrm{c}(a_0)} &= \frac{6144}{425}\sqrt{2}\epsilon^{7/2} + \displaystyle\sum\limits_{i=1}^{\infty}\eta_{i}\epsilon^ {7/2 + i/2} \label{eqn:high_limit_series} \\
   & \!\begin{multlined}[b][0.723\linewidth] \approx\frac{6144}{425}\sqrt{2}\epsilon^{7/2} - 42.0833...\, \epsilon^{4}{} \\ +50.2334...\, \epsilon^{9/2} + ...\: .\label{eqn:high_limit_series_approx}
    \end{multlined}
\end{align}
The integral in Eq.~\eqref{eqn:peters} cannot be solved with standard functions but it can be written in terms of Appell hypergeometric functions. Therefore, the full elliptical merging time (Eq.~\ref{eqn:peters}) requires the calculation of a numerical integral. We use the form of the low- and high-eccentricity limits (Eq.~\ref{eqn:high_limit}--\ref{eqn:high_limit_series_approx}) to infer a numerical functional fit to the elliptical merging time to improve computational efficiency. Our numerical fit is
\begin{align}
    \frac{T(a_0,e_0)}{T_\mathrm{c}(a_0)} = \Bigg[\frac{768}{425} + \displaystyle\sum\limits_{i=1}^{10}\beta_{i}\big(1 - e_{0}^2 \big) ^ {i/2} \Bigg] \big(1 - e_{0}^2 \big) ^{7/2}\: , \label{eqn:numerical_fit}
\end{align}%
\begin{figure*}
\setcounter{section}{3} 
\setcounter{figure}{0} 
  \centering
  \begin{subfigure}{0.34\linewidth}
    \centering
    \includegraphics[width=\linewidth]{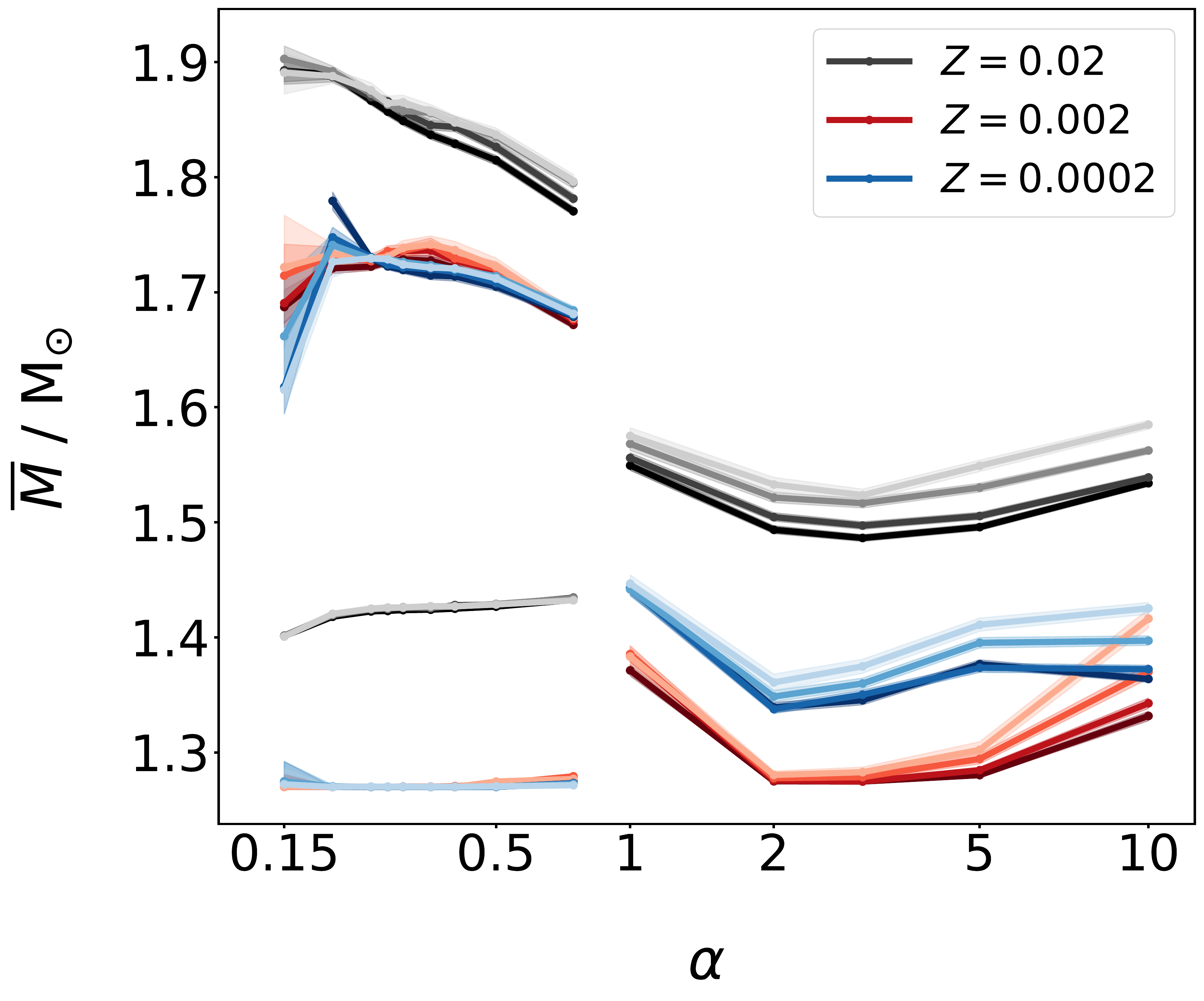}
    \caption{Mean individual mass components.}\label{fig:individual_mass_split}
  \end{subfigure}
    \hspace{15ex}
  \begin{subfigure}{0.34\linewidth}
    \centering
    \includegraphics[width=\linewidth]{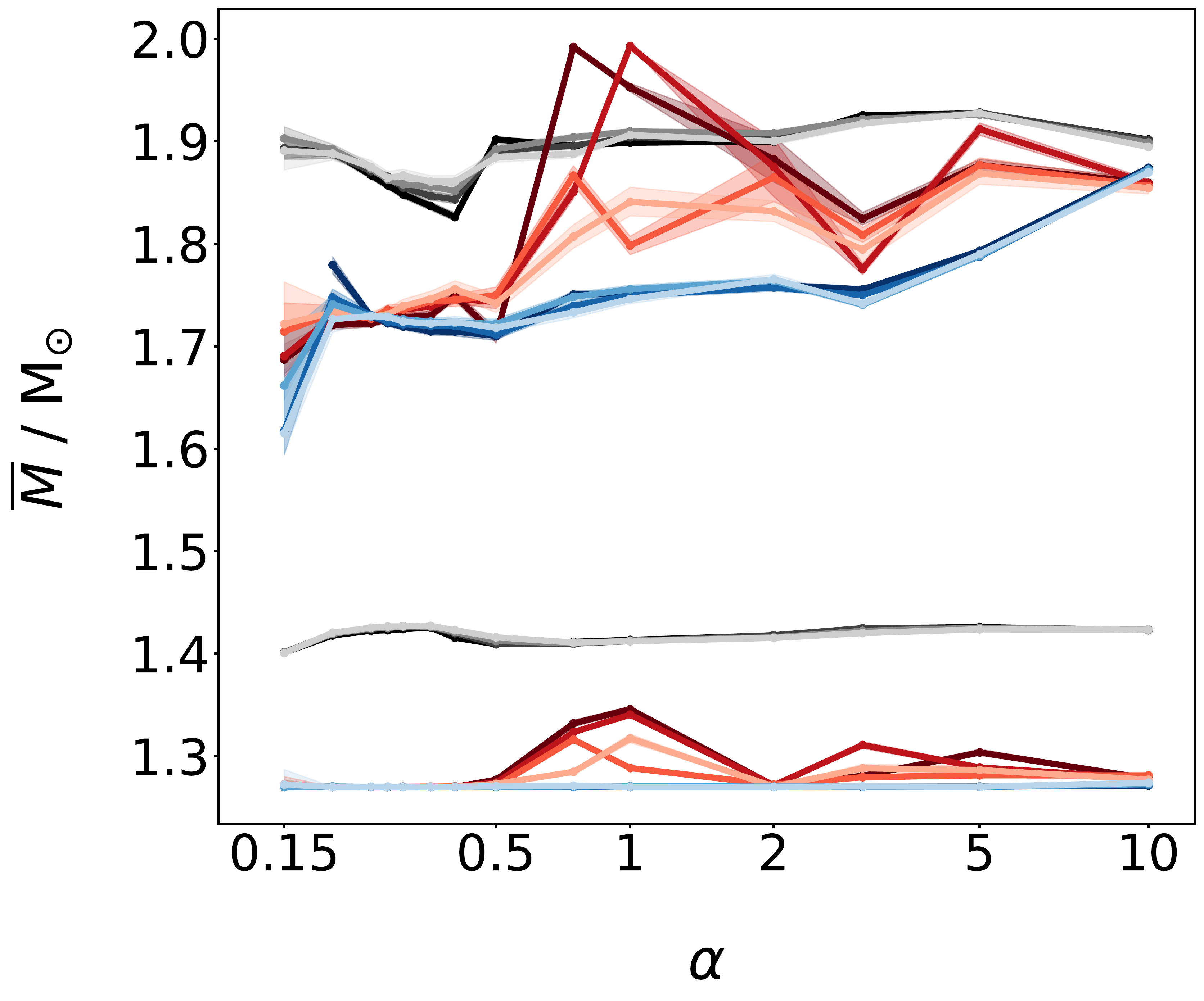}
    \caption{Mean individual merging mass components.}\label{fig:individual_mass_merge_split}
  \end{subfigure}\\  
  \caption{A comparison of the mean individual NS (comprising a DNS) mass~$\overline{M}$ (with bootstrapped $95$ per cent confidence intervals) as functions of~$\alpha$ in models with ECSN kick speed dispersions of~$\sigma_\mathrm{ECSN} \in \{0, 7, 15, 26\} \, \textnormal{km}\,{\textnormal{s}}^{-1}$ at metallicities of~$Z \in \{0.02, 0.002, 0.0002\}$. In Panel~\subref{fig:individual_mass_merge_split} DNSs are considered to be capable of merging if their delay times are less than~$13.8 \, \mathrm{Gyr}$. Panels~\subref{fig:individual_mass_split} and~\subref{fig:individual_mass_merge_split} show the high- and low-mass component means of the individual NS mass probability distributions when they are non-overlapping bimodal distributions with a distinct separation-point. Otherwise the overall mean is plotted when the distributions are overlapping bimodal or unimodal. In the plots the colours are sequential with brighter colours representing higher ECSN kick dispersions and different hues representing different metallicities.} 
  \label{fig:mass_stats_individual}
\setcounter{section}{1} 
\end{figure*}%
where the constants are given in Table~\ref{tab:constants}. In Eq.~\eqref{eqn:peters} $T / T_\mathrm{c}$~is very much less than one for large eccentricities so the fractional errors of any numerical fit are more sensitive to absolute errors at high eccentricities. Consequently, the numerical fit (Eq.~\ref{eqn:numerical_fit}) has been chosen to be consistent with the high-eccentricity limit to leading order exactly (cf. Eq.~\ref{eqn:high_limit}) and to the next two orders approximately (Eq.~\ref{eqn:high_limit_series_approx}) by appropriately choosing the coefficients~$\beta_1$ and~$\beta_2$. Coefficients~$\beta_i$ for~$i \in \{3,4,...,10\}$ were fitted to Eq.~\eqref{eqn:peters} with a least-squares method. The form of the numerical fit has been crafted to ensure that the exponents of its series expansion about~$e_{0} = 0$ and~$e_{0} = 1$ have the same form as those of the elliptic equation. These are~$2i$ and~$(7 + i)/2$, respectively (Eq.~\ref{eqn:low_limit_series} and~\ref{eqn:high_limit_series}).

The numerical fit for~$T / T_\mathrm{c}$ (Eq.~\ref{eqn:numerical_fit}) is plotted against the full elliptical ratio (Eq.~\ref{eqn:peters}), as well as its low- and high-eccentricity limits (Eq.~\ref{eqn:low_limit} and~\ref{eqn:high_limit}), in Fig.~\ref{fig:peters_comparison_linear}. The absolute error between \citeauthor{GW_rad}' (\citeyear{GW_rad}) ratio and our numerical fit
\begin{align}
    \varepsilon_\mathrm{abs} &\equiv \bigg|\frac{T_{\text{Peters}} - T_{\text{Numerical}}}{T_\mathrm{c}} \bigg| \: , \label{eqn:absolute_error}
\end{align}
and the fractional error
\begin{align}
   \varepsilon_\mathrm{frac} &\equiv \bigg| \frac{T_{\text{Peters}} - T_{\text{Numerical}}}{T_{\text{Peters}}} \bigg| \: , \label{eqn:fractional_error}
\end{align}
are displayed in Fig.~\ref{fig:peters_fit_errors} and~\ref{fig:peters_fit_frac_errors}. The fractional errors in our numerical fit have an upper bound of about $\varepsilon_\mathrm{frac} \approx 2 \times 10^{-7}$.

\section{Probability of events and expectations over the grid}
\label{sec:appendixb}

Let the set of tuples of initial grid parameters~$\widetilde{\Omega}$ be given by the Cartesian product of the individual sets of initial grid parameters~$\widetilde{\Omega}_X$ we are considering,
\begin{align}
     \widetilde{\Omega} \equiv \prod_{X \in J} \widetilde{\Omega}_X \, ,
\end{align}
where $J$~is an index set whose elements are the continuous random variables from Eq.~\eqref{eqn:pdf_MqP}. $J$~will either be~$\{M, Q , P\}$ or~$\{M, Q, P, \log Z\}$ depending on whether or not we are implementing metallicity evolution. Necessarily, $\widetilde{\Omega}_X$~are discrete subsets of the continuous sample spaces~$\Omega_X$ from Eq.~\eqref{eqn:pdf_MqP}. The discrete sample spaces~$\widetilde{\Omega}_X$ can be used to construct a partition~$\{I_{\omega_X} : \omega_X \in \widetilde{\Omega}_X\}$ of the continuous sample spaces~$\Omega_X$ such that 
\begin{align}
     \Omega_X = \bigcup_{\omega_X \in \widetilde{\Omega}_X} \!\!\! I_{\omega_X} \, ,
\end{align}
where $I_{\omega_X}$ are disjoint intervals centred on $\omega_X$.

The expectation of some random variable~$G : \Omega \rightarrow \mathbb{R}$ discretely sampled over~$\widetilde{\Omega}$ is given by 
\begin{align}
    \mathbb{E}[G] = \sum_{\omega \in \widetilde{\Omega} } G(\omega) \prod_{ X \in J} \mathbb{P}_{X} \big( X(I_{\omega_X}) \big)  \, , \label{eqn:expectation}
\end{align}
where $\mathbb{P}_{X}(S) \equiv \int_{S} \varphi_X(x) \, \mathrm{d} x$ are the probability measures associated with the probability densities~$\varphi_X$ from Eq.~\eqref{eqn:pdf_MqP} and~\eqref{eqn:pdf_Z}. Similarly, the probability measure of some countable event~$A \subseteq \widetilde{\Omega}$ (e.g. the set of DNSs) in the simulation grid is given by
\begin{align}
    \mathbb{P}(a \in A) \equiv \mathbb{E}[ \mathbf{1}_A] \, , \label{eqn:probability}
\end{align}
where $\mathbf{1}$ is the indicator function.

\section{NS masses}
\label{sec:appendixc}

Fig.~\ref{fig:individual_mass_split} shows the mass distributions of the individual NSs in DNSs. At~$\alpha < 1.0$ these are non-overlapping bimodal distributions and, as~$\alpha$ increases, the components of the bimodal distributions start to overlap. At high~$\alpha$ ($\geq 3.0$) the low-mass component grows in height dominating the high-mass component leading to an effectively unimodal distribution (Fig.~\ref{fig:mass_stacked_hist}). In the non-overlapping bimodal regime, the high-mass component mean individual NS mass varies between~$1.615_{-0.017}^{+0.034}$ and~$1.903_{-0.012}^{+0.011} \, \mathrm{M}_\odot$ and the low-mass component mean varies between~$1.27$ and~$1.4345_{-0.0014}^{+0.0014} \, \mathrm{M}_\odot$ (Fig.~\ref{fig:individual_mass_split}) depending on~$\alpha$,~$\sigma_\mathrm{ECSN}$ and~$Z$. The low-mass component mass distributions are symmetric with small dispersions of up to~$0.0391_{-0.0013}^{+0.0013} \, \mathrm{M}_\odot$ whilst the high-mass component mass distributions are often skewed with larger dispersions up to~$0.110_{-0.004}^{+0.003} \, \mathrm{M}_\odot$.

Our ranges of mass distribution summary statistics are consistent with those of \citet{bimodalnsmass2018}, who found strong evidence for a bimodal distribution of masses of NSs in DNSs. They inferred a narrow low-mass component of~$1.34_{-0.02}^{+0.03} \, \mathrm{M}_\odot$ with a dispersion of~$0.07_{-0.02}^{+0.02} \, \mathrm{M}_\odot$ and a wider high-mass component of~$1.80_{-0.18}^{+0.15} \, \mathrm{M}_\odot$ with a dispersion of~$0.21_{-0.14}^{+0.18} \, \mathrm{M}_\odot$. None of the mean mass components at a single metallicity on their own are consistent with these observations. This may suggest that NSs form from progenitors with a range of metallicities.

The merging individual NS mass distributions remain non-overlapping bimodal distributions at all~$\alpha$ (Fig.~\ref{fig:individual_mass_merge_split}). By comparing the mean total and mean individual merging mass it is possible to infer which combinations of low- and high-mass component mass NSs are most common in merging DNSs. Fig.~\ref{fig:total_mass_merge} and Fig.~\ref{fig:individual_mass_merge_split} suggest that most frequently a merging DNS consists of one low-mass and one high-mass NS. The exceptions are for~$Z = 0.002$ at~$\alpha = 2.0$ and for~$Z = 0.0002$ at ~$\alpha \leq 0.20$, for which the most frequent composition is two low-mass NSs.

Although the mean total merging DNS mass is greater than or equal to the mean total overall DNS mass, the mean merging individual NS masses are not always greater than or equal to the mean individual overall NS masses. The high-mass component mean merging individual NS mass varies between~$1.615_{-0.017}^{+0.030}$ and~$1.99 \, \mathrm{M}_\odot$ and the low-mass component mean varies between~$1.27$ and~$1.4272_{-0.0021}^{+0.0020}\, \mathrm{M}_\odot$ (Fig.~\ref{fig:individual_mass_merge_split}) depending on~$\alpha$,~$\sigma_\mathrm{ECSN}$ and~$Z$. The low-mass component mass distributions are symmetric with dispersions of up to $0.0821_{-0.0011}^{+0.0010} \, \mathrm{M}_\odot$ whilst the high-mass component mass distributions are often skewed with larger dispersions up to~$0.167_{-0.019}^{+0.015} \, \mathrm{M}_\odot$ (Fig.~\ref{fig:mass_stacked_hist}).

Our mass data are simulations of the astrophysical DNS mass distributions. Measurements of radio pulsars and gravitational waves are affected by selection effects. Therefore, when comparing our data with that of studies of radio pulsars or gravitational wave sources it is important to compensate for selection effects to ensure a fair comparison. We used some studies that already consider these effects \citep{bimodalnsmass2018, review_selection_effects_chattopadhyay, LIGO_BNS_merger_rate_2021, mass_GW_inference_landry}.

\begin{figure*}
\setcounter{figure}{1} 
  \centering
    \begin{subfigure}{.95\linewidth} 
    \centering
    \includegraphics[width=\linewidth]{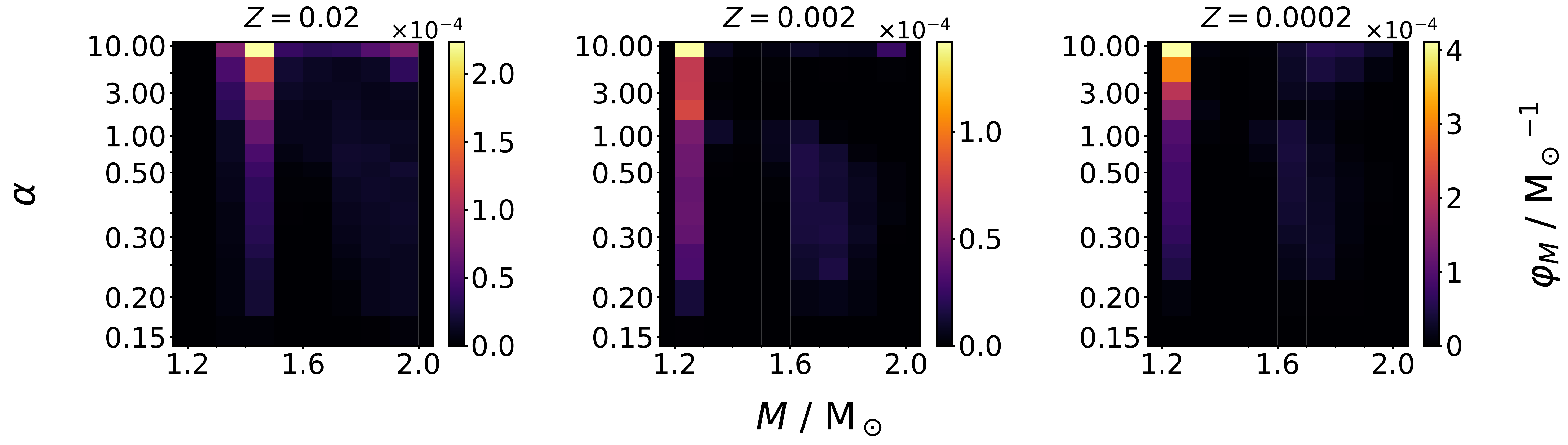}
    \caption{Individual NS masses that constitute DNSs.}
  \end{subfigure}\\
  \vspace{1ex}
    \begin{subfigure}{.95\linewidth} 
    \centering
    \includegraphics[width=\linewidth]{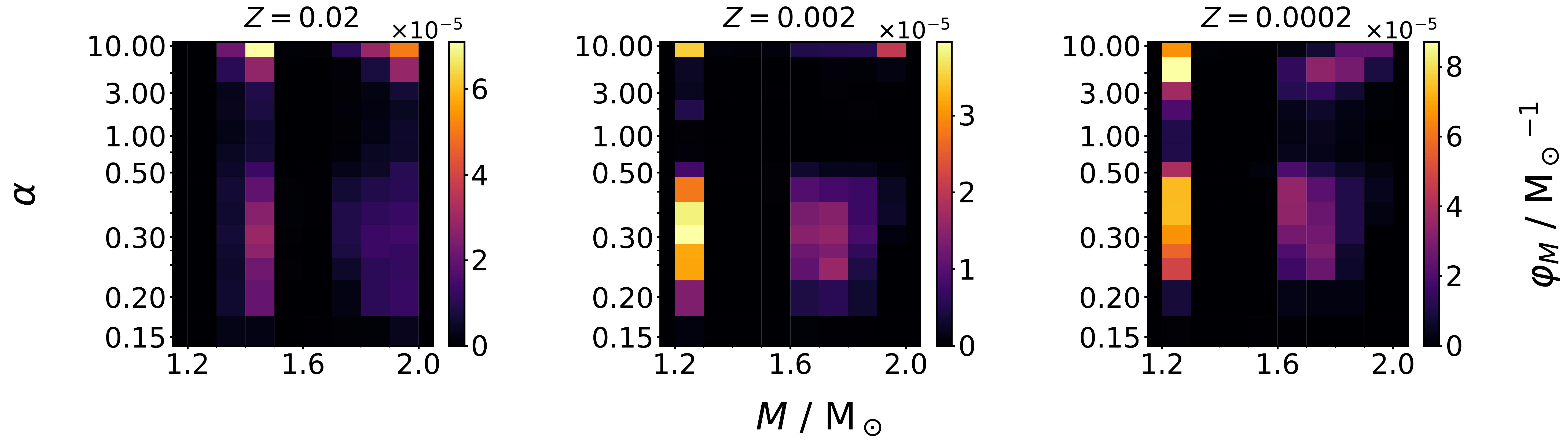}
    \caption{Individual NS masses that constitute merging DNSs.}
  \end{subfigure}\\
  \vspace{1ex}
    \begin{subfigure}{.95\linewidth} 
    \centering
    \includegraphics[width=\linewidth]{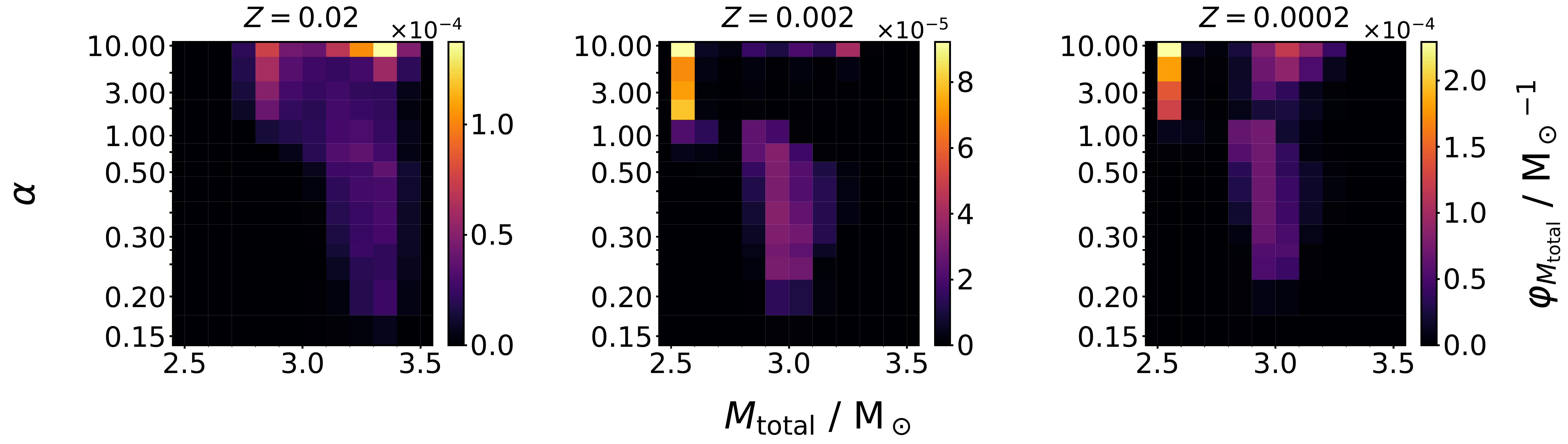}
    \caption{Total DNS masses.}
  \end{subfigure}\\
  \vspace{1ex}
    \begin{subfigure}{.95\linewidth} 
    \centering
    \includegraphics[width=\linewidth]{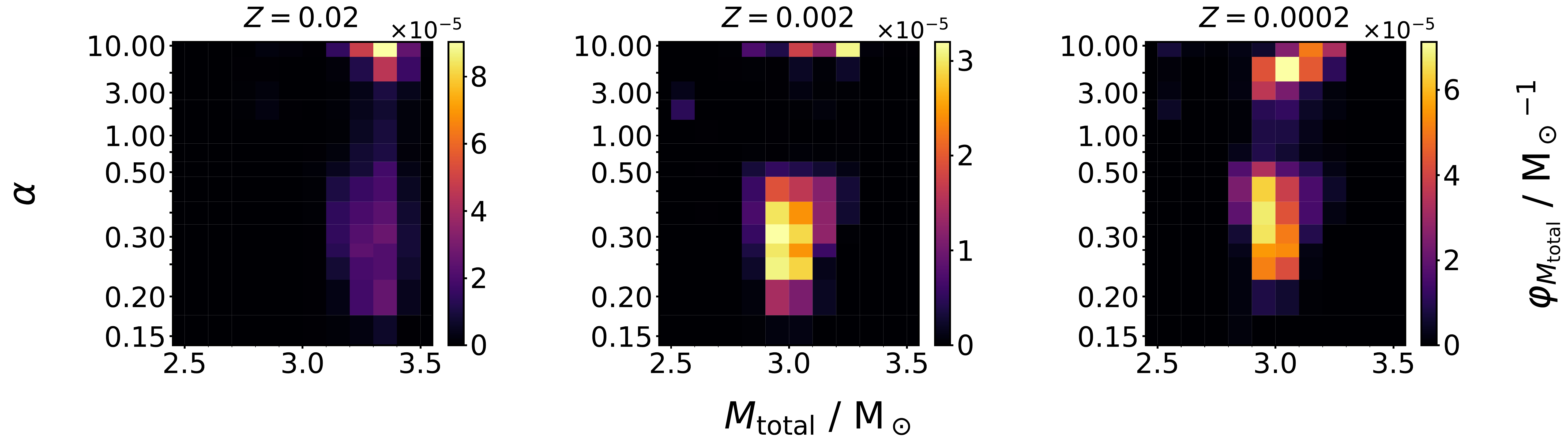}
    \caption{Total merging DNS masses.}
  \end{subfigure}\\
  \caption{Stacked histograms, for each~$\alpha$, of the total DNS mass~$M_\mathrm{total}$ and individual NS mass~$M$ for models with ECSN kick dispersion of~$\sigma_\mathrm{ECSN} = 26\, \textnormal{km}\,{\textnormal{s}}^{-1}$, where~$\varphi _{M}$ and~$\varphi _{M_\text{total}}$ are the probability densities.}
  \label{fig:mass_stacked_hist}
\end{figure*} 

\section{DNS period-eccentricity distributions}
\label{sec:appendixd}

\begin{figure*}
    \centering    \includegraphics[width=0.93\linewidth]{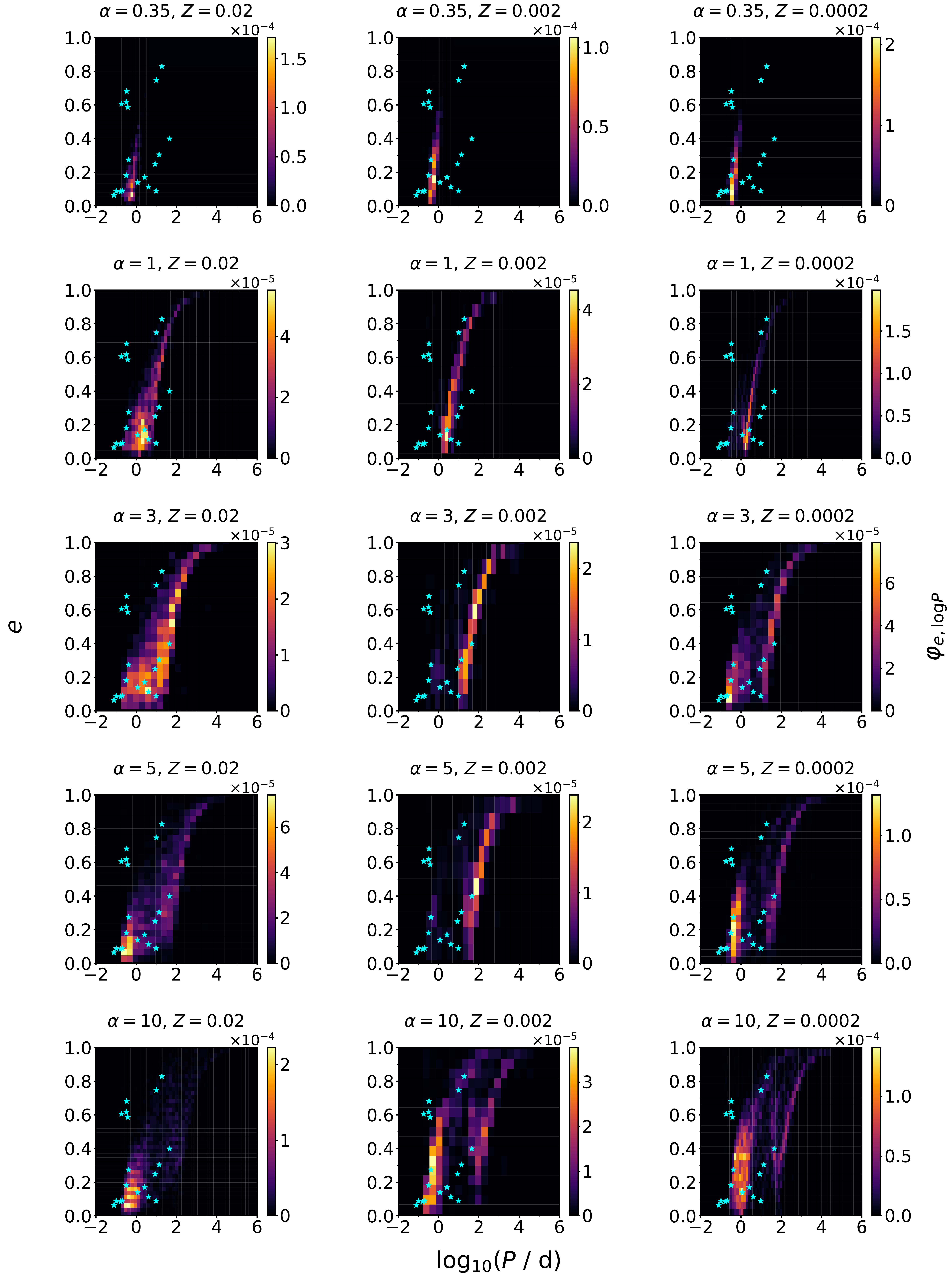}
  \caption{Period-eccentricity histograms of DNS systems upon formation for models at various $\alpha$ values and metallicities $Z$ with an ECSN kick dispersion of~$\sigma_\mathrm{ECSN} = 26\, \textnormal{km}\,{\textnormal{s}}^{-1}$, where~$\varphi_{e,\log P}$ are the probability densities.  Observations of Galactic binary pulsars are plotted on top of our histograms.}
  \label{fig:p_e_dists}
\end{figure*}

Fig.~\ref{fig:p_e_dists} shows the period-eccentricity distributions of DNS systems upon formation in our simulations at five different CE ejection efficiencies $\alpha \in \{0.35, 1, 3, 5, 10\}$, for~$\sigma_\mathrm{ECSN}=26\,\mathrm{km}\,\mathrm{s}^{-1}$ and $Z \in \{0.02, 0.002, 0.0002\}$. Observed orbital parameters of Galactic pulsars are shown for comparison, with the observed parameters taken from \cite{P_e_Champion2004}, \cite{P_e_Faulkner2005}, \cite{P_e_Jacoby2006}, \cite{P_e_Kramer2006}, \cite{P_e_Lorimer2006}, \cite{P_e_Corongiu2007}, \cite{P_e_Janssen2008}, \cite{P_e_Keith2009}, \cite*{P_e_Weisberg2010}, \cite{P_e_Lynch2012_possibly_a_WD, P_e_Lynch2018}, \cite*{P_e_Fonseca2014}, \cite{P_e_Swiggum2015}, \cite{P_e_Martinez2015}, \cite{P_e_Ng2015_possibly_a_WD}, \cite{P_e_Martinez2017}, \cite{P_e_Lazarus2016}, \cite{P_e_Cameron2018} and \cite{P_e_Stovall2018}. We find that our simulations qualitatively fit the observed properties better as~$\alpha$ increases. Similar comparisons, using other codes, were performed by \cite{Tauris2017}, \cite{comparison_kruckow_2018} and \cite{VignaGomez2018}.

\bsp 
\label{lastpage}

\end{document}